\documentclass[acmsmall,screen]{acmart}

\usepackage{enumerate}
\usepackage{ragged2e}

\usepackage{stfloats}
\usepackage{amsmath,amsfonts}
\usepackage{algorithmic}
\usepackage{graphicx}
\usepackage{textcomp}
\usepackage{xcolor}
\usepackage{multirow}
\usepackage{color}
\usepackage{subfigure}
\usepackage{diagbox}

\usepackage{url}
\usepackage{hyperref}
\hypersetup{
    colorlinks=true,
    linkcolor=blue,
    filecolor=blue,      
    urlcolor=blue,
    citecolor=blue,
}

\usepackage{multirow}
\usepackage[utf8]{inputenc}
\usepackage{algorithmic}
\usepackage{algorithm}
\usepackage{graphicx}
\usepackage{textcomp}
\usepackage{color,xcolor}
\usepackage{url}
\usepackage{graphicx}
\usepackage{subfigure}
\usepackage{stmaryrd}
\usepackage{verbatimbox}
\usepackage{enumerate}
\usepackage[shortlabels]{enumitem}
\usepackage{bm}

\usepackage{caption}

\usepackage{makecell}
\usepackage{booktabs}

\usepackage{multirow}

\usepackage{enumitem}
\usepackage{colortbl}
\definecolor{codegreen}{rgb}{0,0.6,0}
\definecolor{codegray}{rgb}{0.5,0.5,0.5}
\definecolor{codepurple}{rgb}{0.58,0,0.82}
\definecolor{backcolour}{rgb}{0.95,0.95,0.92}

\usepackage{listings}
\lstdefinestyle{mystyle}{
  backgroundcolor=\color{backcolour},   commentstyle=\color{codegreen},
  keywordstyle=\color{magenta},
  numberstyle=\tiny\color{codegray},
  stringstyle=\color{codepurple},
  basicstyle=\ttfamily\footnotesize,
  breakatwhitespace=false,
  breaklines=true,
  captionpos=b,
  keepspaces=true,
  numbers=left,
  numbersep=5pt,
  showspaces=false,
  showstringspaces=false,
  showtabs=false,
  tabsize=2
}
\usepackage[most]{tcolorbox}
\newcommand{\myfinding}[2]{
\begin{center}
\begin{tcolorbox}[colback=gray!15, colframe=black, boxsep= -0.15cm, middle=-0.15cm, breakable]
\textbf{Answer to RQ{#1}:}
{#2}
\end{tcolorbox}
\end{center}
}

\usepackage{blindtext}
\newenvironment{mylist}[1]%
{\begin{list}{}{\settowidth{\labelwidth}{\bf #1}%
			\setlength{\leftmargin}{\labelwidth}%
			\addtolength{\leftmargin}{\labelsep}%
			}}%
{\end{list}}

\usepackage{xspace}
\newcommand{\ie}{\textit{i.e.,}\xspace}
\newcommand{\eg}{\textit{e.g.,}\xspace}

\newcommand{\etal}{\textit{et al.}\xspace}

\newcommand{\figref}[1]{Fig.~\ref{#1}\xspace}

\usepackage[normalem]{ulem}

\begin{document}


\title{On the Effectiveness of Code Representation in Deep Learning-Based Automated Patch Correctness Assessment}


\author{Quanjun Zhang}
\email{quanjunzhang@njust.edu.cn}
\orcid{0000-0002-2495-3805}
\affiliation{
\institution{Nanjing University of Science and Technology}
\city{Nanjing}\country{China}
}

\author{Haichuan Hu}
\orcid{0009-0002-3007-488X}
\email{181250046@smail.nju.edu.cn}

\author{Chunrong Fang}
\email{fangchunrong@nju.edu.cn}
\orcid{0000-0002-9930-7111}

\author{Ye Shang}
\email{yeshang@smail.nju.edu.cn}
\orcid{0009-0000-8699-8075}

\author{Tao Zheng}
\email{zt@nju.edu.cn}
\orcid{0}

\author{Zhenyu Chen}
\email{zychen@nju.edu.cn}
\orcid{0000-0002-9592-7022}

\affiliation{
  \institution{Nanjing University}
  \city{Nanjing}
  \state{Jiangsu}
  \country{China}
}

\author{Yun Yang}
\email{yyang@swin.edu.au}
\orcid{0000-0002-7868-5471}
\affiliation{
\institution{Swinburne University of Technology}
\city{Melbourne}\country{Australia}
}

\author{Liang Xiao}
\email{xiaoliang@mail.njust.edu.cn}
\orcid{0000-0003-0178-9384}
\affiliation{
\institution{Nanjing University of Science and Technology}
\city{Nanjing}\country{China}
}

\begin{abstract}

Automated program repair (APR) attempts to generate correct patches and has drawn wide attention from both academia and industry in the past decades.
However, APR is continually struggling with the patch overfitting issue due to the weak test suites, \ie generating plausible but overfitting patches.
Thus, to address the overfitting problem, the community has proposed an increasing number of automated patch correctness assessment (APCA) approaches.
Among them, leveraging deep learning approaches to predict patch correctness has been emerging recently.
Such approaches typically encode patched code snippets into well-designed representations (\eg hand-crafted features) and build a binary model for correctness prediction.
Despite being fundamental in reasoning about patch correctness, code representation has not been systematically investigated, and little is known about its advantages and disadvantages.



To bridge this gap, we perform the first extensive study to evaluate the performance of different code representations on predicting patch correctness, involving more than 500 trained APCA models.
The experimental results on 15 representations from four categories and 11 classifiers show that the graph-based representation, which is ill-explored in the literature, consistently outperforms other representations, \eg an average accuracy of 82.69\% for CPG across three GNN models.
Moreover, we demonstrate that such representations can achieve comparable or better performance against previous APCA approaches, \eg filtering out 87.09\% overfitting patches by TreeLSTM trained with AST.  
We further explore the capacity and limitation of fusing different types of representation, and find that integrating sequence-based representation into heuristic-based representation is able to yield an average improvement of 13.58\% on five metrics.
Besides, we provide additional discussion about the node embedding in the well-performing graph-based representation, and illustrate that the node textual information is more critical than the node type information in learning patch semantics.
Inspired by the findings, we finally pinpoint practical guidelines for advanced representation-based APCA studies, such as combining more representations.
Overall, our study highlights the potential and challenges of utilizing code representation to reason about patch correctness, thus increasing the usability of off-the-shelf APR tools and reducing the manual debugging effort of developers in practice.

\end{abstract}

\ccsdesc[500]{Software and its engineering~Software testing and debugging}
\keywords{Automated Program Repair, Patch Overfitting, Code Representation}

\maketitle

\section{Introduction}
\label{sec:intro}

During the maintenance and evolution of modern software systems, software bugs are inevitable and may result in fatal consequences~\cite{gazzola2019automatic}.
It is increasingly time-consuming and labor-intensive for developers to manually fix detected bugs due to the rising number and complexity of software systems~\cite{hossain2018challenges}.
For example, prior work~\cite{britton2013reversible} demonstrates that developers usually spend 50\% of their time on software debugging and fixing.
To relieve manual debugging efforts, \textbf{Automated Program Repair (APR)} has been proposed to generate correct patches automatically without human intervention, and has received a great deal of attention from both academia and industry~\cite{zhang2023survey,zhang2024systematic}.
In the past decades, a mass of APR approaches have been proposed and have made substantial progress on the number of correctly-fixed bugs, including heuristic-based, constraint-based, pattern-based, and learning-based ones~\cite{monperrus2018automatic,le2012genprog,liu2019tbar,li2020dlfix,xia2025demystifying}.


However, the APR community has fundamentally been challenged by the \textit{patch overfitting issue} due to insufficient program specifications~\cite{smith2015cure,le2019reliability,liang2021interactive,motwani2022quality,motwani2023better}.
In particular, existing APR approaches typically adopted the \textit{generate-and-validate} paradigm, \ie generating candidate patches and leveraging the developer-written test cases to identify correct patches~\cite{zhang2023survey}.
In such cases, candidate patches passing the available test cases are returned, but they may not be generalized to other potential test cases, leading to overfitting patches.
Previous studies~\cite{wang2025show,tao2014automatically} demonstrate that it may take considerable time and effort for developers to filter out the overfitting patches, hindering the practical usability of off-the-shelf APR tools in real-world debugging scenarios.

Thus, to address the patch overfitting issue, various \textbf{Automated Patch Correctness Assessment (APCA)} approaches have been proposed to determine whether a plausible patch is indeed correct or not~\cite{tan2016anti,xiong2018identifying,xin2017difftgen,yang2020daikon,le2017s3,dong2024method,ismayilzada2023poracle}.
Traditional APCA techniques can be categorized into two groups according to extracted features, including static and dynamic ones.
Static approaches attempt to analyze code-changed patterns or code similarity based on syntactic and semantic features.
For example, Tan~\etal~\cite{tan2016anti} define a set of generic forbidden transformations (such as deleting functionality) for the buggy program. 
On the contrary, dynamic approaches usually execute candidate patches against additional test cases generated by automated test generation tools.
For example, Xiong~\etal~\cite{xiong2018identifying} generate new test cases and identify overfitting patches based on the execution trace similarity.
However, static techniques may face challenges in achieving high identification precision, while dynamic techniques require intensive time to collect runtime information~\cite{wang2020automated}.

Recently, inspired by the advance of \textbf{Deep Learning (DL)}, an increasing number of learning-based APCA techniques have been proposed to directly predict patch correctness~\cite{ye2021automated,tian2020evaluating,tian2022change,tian2022best,tian2022predicting,lin2022context,zhang2024appt,le2023invalidator,ghanbari2022patch}.
Different from traditional APCA, learning-based APCA is able to learn the relationships of correct and overfitting patches from corpora of patch benchmarks, thus receiving rapidly increasing attention~\cite{wang2020automated,yang2023large}.
Such techniques typically handle the APCA problem as a code classification task with a two-fold prediction pipeline, \ie a feature extractor and a binary classifier.
The \textit{feature extractor} first extracts the hidden features of patched code snippets, and the \textit{binary classifier} takes the extracted features to perform patch correctness prediction.
For example, He~\etal~\cite{ye2021automated} extract hand-crafted features from Java programs statically and train a probabilistic model to perform patch correctness prediction.
Besides, Tian~\etal~\cite{tian2020evaluating} investigate the feasibility of embedding features to build predictive models by representing source code as a sequence of tokens.
Furthermore, Lin~\etal~\cite{lin2022context} utilize abstract syntax trees (ASTs) to represent patched code snippets for patch classifier training.
Thanks to the well-constructed code representations and powerful DL models, learning-based APCA techniques have achieved promising performance in the last couple of years~\cite{zhang2023survey,yang2023large}.

\textbf{This Paper.}
Despite an emerging research area, the APCA community has seen different well-designed code representation approaches to encode patched code snippets, \eg heuristic-based~\cite{ye2021automated}, sequence-based~\cite{tian2020evaluating}, and tree-based features~\cite{lin2022context}.
Code representation is crucial and fundamental in learning the program semantics and reasoning about patch correctness, and has already been widely investigated in various code understanding tasks~\cite{siow2022learning,sun2023abstract,namavar2022controlled,utkin2022evaluating}.
However, the literature does not yet provide a systematic evaluation of how to represent source code for the patch correctness assessment task.
There exist several unsolved challenges about code representations in APCA.
(1) What type of code representation is optimal in the patch correctness assessment scenario?
(2) Despite the superiority shown by heuristic-based, sequence-based, and tree-based approaches, the graph-based approach remains underexplored.
(3) Can such different representations be integrated to further enhance the prediction effectiveness?
In this paper, to fill this gap, we investigate the feasibility of leveraging different code representation approaches to predict patch correctness and facilitate subsequent APCA studies.
Based on the above challenges, we address the following three research questions (RQs):
RQ1 analyzes the impact of different representations; RQ2 answers the comparison performance with state-of-the-art approaches; and RQ3 explores the potential of fusion-based approaches.

\begin{mylist}{\textbf{(RQ1)}}
  \item[\textbf{(RQ1)}] \textbf{The impact of different code representation approaches in reasoning about patch correctness.}
  
  \textbf{\underline{Results:}}
  Among four types of representations, graph-based representation exhibits the best performance, tree-based and sequence-based representations achieve comparable results, with heuristic-based representation performing slightly less effectively.
  For example, as the best-performing approach within each type of representation, XGBoost with TF-IDF, Transformer with sequence, TreeLSTM with AST and GGNN with CPG achieve an accuracy of 80.41\%, 82.48\%, 82.94\%, and 83.73\%, respectively.

  \item[\textbf{(RQ2)}] \textbf{The comparison of four types of code representation with state-of-the-art APCA approaches.}
  
  \textbf{\underline{Results:} }
  Four types of representation achieve an average performance of 80.55\%, 82.90\%, 83.03\%, and 83.81\% across all investigated metrics, which are comparable to or exceed those of previous APCA approaches.
  For example, the graph-based representation (CPG+GGNN) outperforms Tian~\etal~\cite{tian2020evaluating} (BERT+SVM) by 9.34\%, 14.96\% and 8.83\% on accuracy, recall and F1, respectively, demonstrating the substantial benefits of code representation in APCA.

  \item[\textbf{(RQ3)}] \textbf{The  feasibility of fusion-based APCA approaches.}
  
  \textbf{\underline{Results:} }
  It is beneficial to combine two representations, such as an average improvement of 13.58\% brought by integrating sequence-based representation into the heuristic-based representation, indicating the potential ability of feature fusion in APCA.
  While it is challenging to combine more than two representations, such as an average decrease of 3.34\% brought by integrating tree-based representation into heuristic-and-sequence-based representation, indicating in-depth future research efforts.
  
\end{mylist}

We also provide extended discussion to demonstrate that, for the graph-based representation, textual node information yields better results than type node information due to the rich code semantics in reasoning about patch correctness.
Finally, based on our findings, we pinpoint open research challenges and provide practical guidelines for investigating code representation in future APCA studies.
\looseness=-1

\textbf{Novelty \& Contributions.}
To sum up, the main contributions of this paper are as follows:

\begin{itemize}

  
    \item \textbf{Patch benchmark.}
    We construct a high-quality, large-scale patch benchmark to evaluate APCA approaches, with 2,274 labeled plausible patches from real-world Defects4J bugs generated by more than 30 repair tools. 

    \item \textbf{Extensive Study.} 
    We perform a large-scale empirical study to evaluate different code representation approaches in the crucial APCA task.
    Our work involves four types of representations across 20 representation-classifier scenarios and more than 500 trained models.

    \item \textbf{Representation Fusion.} 
    We explore the potential ability to combine different types of representation in reasoning about patch correctness, involving 22 fusion scenarios across 11 combined representations and two fusion strategies, \ie intermediate and backend fusion.

    \item \textbf{Practical Guidelines.}
    We provide practical guidelines on applying code representations for future APCA research, such as exploring advanced fusion approaches.

\end{itemize}

\textbf{Open Science.}
To facilitate future research in the APCA community, we release the replication artifacts, including the studied dataset, source code, model training and evaluation scripts, and related models in our experiment~\cite{myurl}.


\textbf{Paper Organization.} 
Section~\ref{sec:bg&mv} reviews background information.
Section~\ref{sec:study} introduces our empirical study design.
Section~\ref{sec:re&an} analyzes detailed results and answers three RQs.
Section~\ref{sec:dis} presents additional discussions, and
Section~\ref{sec:related_work} provides related work.
Section~\ref{sec:conclusion} draws the conclusions.

\section{Background}
\label{sec:bg&mv}
\subsection{Automated Program Repair}

\begin{figure}[htbp]
\centering
\graphicspath{{graphs/}}
    \includegraphics[width=0.7\linewidth]{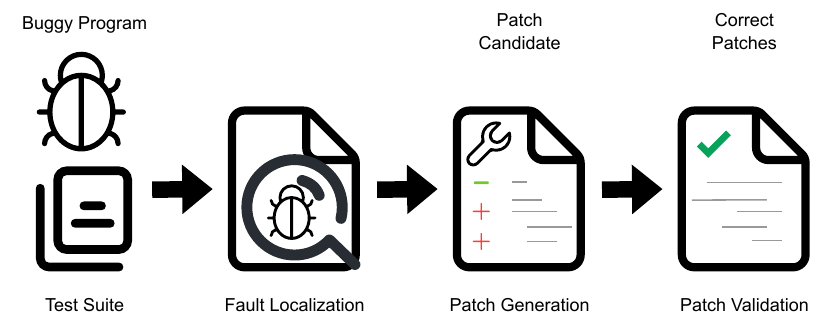}
    \caption{Overview of APR}
    \label{fig:apr}
\end{figure}

\subsubsection{Repair Workflow}
Given a buggy program and some available test cases that make the program fail, the primary objective of APR is to generate patches automatically that pass all available test suites.
Existing APR techniques typically follow the generate-and-validate workflow, as illustrated in \figref{fig:apr}.
In particular, for a detected bug, 
(1) the \textbf{fault localization phase} identifies suspicious code elements and returns the code snippets that need to be patched;
(2) the \textbf{patch generation phase} generates candidate patches by performing program transformation while keeping the program semantically correct;
(3) the \textbf{patch validation phase} filters out incorrect patches by executing available test suites against all generated patches.
In the literature, a mass of APR techniques have been proposed to generate patches from different aspects.
APR techniques can be categorized into four classes, including heuristic-based~\cite{le2012genprog, martinez2016astor,yuan2018arja}, constraint-based~\cite{martinez2018ultra,durieux2016dynamoth, mechtaev2016angelix}, pattern-based~\cite{koyuncu2020fixminer,liu2019avatar,liu2019tbar,le2016history} and learning-based~\cite{tufano2019empirical,zhu2021syntax,ye2024iter} ones.
We detail them in Section~\ref{sec:re_apr}.

\subsubsection{Patch Overfitting Issue}
During patch generation, a candidate patch passing the original test suite is called a \textit{plausible} patch.
A plausible patch, which is also semantically equivalent to the developer patch, denotes a \textit{correct} patch.
However, existing test suites are inherently incomplete to cover the entire behavioral domain of a program.
Thus, some plausible patches passing available test suites may not be generalized to potential test suites, resulting in a critical challenge in APR literature, \ie \textbf{the patch overfitting issue}.
For example, Qi~\etal~\cite{qi2015analysis} find that a majority of plausible patches generated by APR approaches are overfitting the available test suite while actually being incorrect.
When provided to developers, such overfitting patches require extensive manual inspection efforts and lead to negative debugging performance~\cite{tao2014automatically,zhang2022program}, limiting the usability of existing APR approaches in practice.


\subsubsection{Patch Correctness Assessment}

It is fundamentally challenging for existing APR approaches to ensure the correctness of the plausible patches due to the weak test suites.
Thus, the literate has seen various patch correctness assessment techniques to filter out overfitting patches automatically.
Traditional APCA techniques can be categorized into two types according to whether the dynamic program execution is required, \ie static-based and dynamic-based techniques~\cite{tan2016anti,xiong2018identifying}.
In particular, (1) static-based APCA relies on static code features such as code-deleting program transformations, while (2) dynamic-based APCA typically requires runtime information obtained by executing test cases on fixed or patched programs.
In addition, an increasing number of learning-based APCA techniques attempt to predict the correctness of plausible patches by machine or deep learning models~\cite{zhang2023survey}.
Such approaches usually utilize source code features, which are manually designed by professional developers or automatically extracted by off-the-shelf code embedding techniques, to train a classifier for patch correctness classification.
In this work, we attempt to explore the performance of different code representation approaches in learning-based APCA, paving the way for future advancements in this community.

\subsection{Code Representation}
\label{sec:ba_representation}

With the success of DL in the SE community, code representation has been foundational for learning the semantics of source code to facilitate code-related task automation~\cite{watson2022systematic,yang2022survey,zhang2026survey}.
According to previous SE studies~\cite{zhang2023survey,siow2022learning,sun2023abstract}, code representation can be categorized into four types, including heuristic-based, sequence-based, tree-based, and graph-based representations.
\textbf{heuristic-based representation} refers to representing source code as a set of hand-crafted features, such as frequency-inverse document frequency (TF-IDF).
Such features are usually pre-designed by domain experts to identify important aspects of source code based on specific tasks, and utilized in various tasks such as patch correctness assessment~\cite{ye2021automated} and fault localization~\cite{wong2016survey}.
\textbf{Sequence-based representation} treats source code as a sequence of tokens, which is similar to natural language.
A mass of studies employ sequence-based representation to address code-related tasks, such as program repair~\cite{yuan2022circle,zhang2023pre} and patch correctness assessment~\cite{tian2020evaluating,tian2022best,zhang2024appt}, demonstrating that the sequence-based representation is capable of learning semantic and syntactic information of source code.
\textbf{Tree-based representation} attempts further to capture the structure information of source code with AST.
Such representation is widely-used in previous studies, such as program repair~\cite{li2020dlfix,li2022dear} and patch correctness assessment~\cite{lin2022context}.
\textbf{Graph-based representation} further considers some semantic information that is not well-modeled in tree-based representation, such as control flow and data flow.
Researchers have proposed various approaches with graph-based representation for source code, such as program repair~\cite{zhu2021syntax,zhu2023tare} and vulnerability detection\cite{li2021vulnerability,zhou2019devign}.
In this work, we consider all four types of code representations to train classification models for patch correctness prediction.

\section{Study Design}
\label{sec:study}

\begin{figure}
    \centering
    \includegraphics[width=0.9\linewidth]{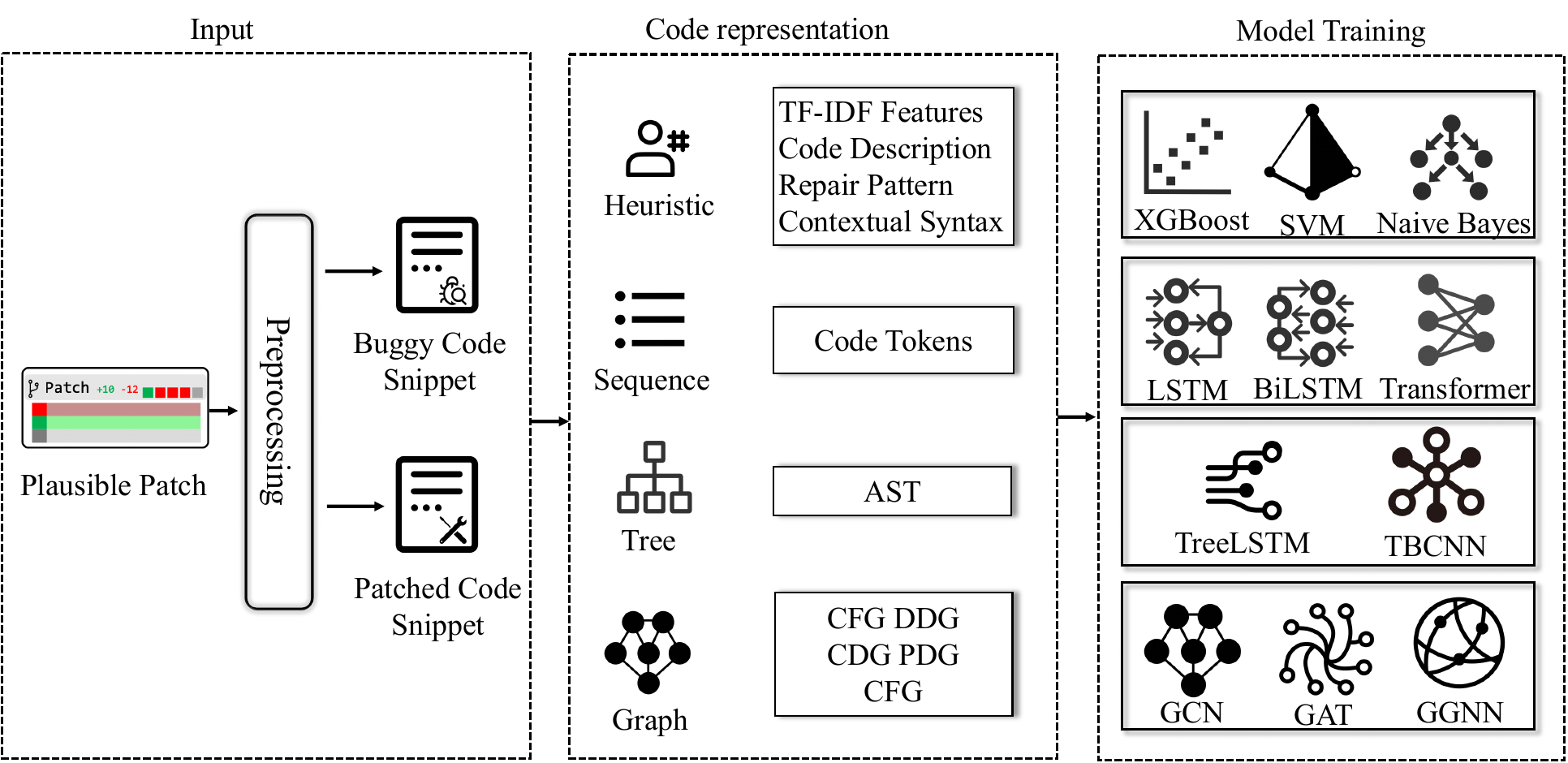}
    \caption{Overview of our study}
    \label{fig:workflow}
\end{figure}

\subsection{Overview}
Overall, our work aims to investigate the effectiveness of various code representations in reasoning about patch correctness.
Fig.~\ref{fig:workflow} presents a typical prediction pipeline of automated patch correctness assessment when DL models are incorporated.
The workflow comprises three primary phases, illustrated as follows.
First, given a plausible patch generated by off-the-shelf APR tools and passing available functional test cases, we perform the input preprocessing phase to extract the buggy code snippet and the patched code snippet.
Second, we represent the buggy and patched code snippet from different perspectives to capture their features, including heuristic-based, sequence-based, tree-based and graph-based strategies.
Third, the extracted code features are used to train DL models to predict patch correctness.

\subsection{Research Questions}

In this work, we explore the following research questions.


\begin{description}
	
  \item [RQ1:] How do four types of code representation approaches perform and compare in predicting patch correctness?
  \item [RQ2:]  What is the performance of four types of representation when compared with state-of-the-art techniques?
  \item [RQ3:] Is it possible to improve the effectiveness of prediction by integrating different types of code representation?
\end{description}

\subsection{Representation Selection}
\label{sec:representation}


Following previous work~\cite{zhang2023survey, siow2022learning}, we consider four major types of code representation strategies in the learning-based SE research, including heuristic-based, sequence-based, tree-based, and graph-based ones, detailed as follows.

\subsubsection{heuristic-based Representation}

We first consider TF-IDF for vectorizing code snippets, which has been widely adopted in prior studies~\cite{siow2022learning}.
In addition, Following ODS~\cite{ye2021automated}, we adopt all 202 hand-crafted features, which are categorized into three types (\ie code, context, and pattern) as per the original design of ODS.



\textbf{TF-IDF Features.}
TF-IDF is a common numerical statistic that reflects the relevance of a code token to the whole code sequence.
In the patch correctness task, it is able to compute a vector for the patched program based on the term frequency and inverse document frequency, thereby diminishing the significance of common stopwords and amplifying the importance of more pertinent code tokens.\looseness=-1

\textbf{Code Features.}
Unlike TF-IDF originally designed for natural language based on token frequency, code features further consider the syntactic and semantic features of code snippets.
We utilize Coming~\cite{martinez2019coming} to consider 50 code features, which can be categorized into four types, including operators, variables, statements, and AST operations.


\textbf{Context Features.}
We consider some contextual features to discover other correlated statements around the buggy statements within the patch, which has been proven to be crucial for predicting patch correctness~\cite{lin2022context}.
These features describe under what context an action happens to an entity and we take all of these as context-affected features.

\textbf{Pattern Features.}
Repair patterns represent common code change actions (\eg insertion of an If statement) and contain valuable semantic information about a given patch.
We follow the taxonomy of repair patterns by Sobreira~\etal~\cite{sobreira2018dissection}, which is built upon 395 Defects4J human-written patches. 
In particular, we utilize ADD~\cite{madeiral2018towards} to automatically extract 26 repair patterns across five categories, \ie wraps-with, expression, conditional, null check, and other patterns.

In addition to the above-mentioned three types of hand-crafted features in ODS~\cite{ye2021automated}, we further explore their expressive power by constructing hybrid feature sets, based on pairwise combinations (\ie code \& context, code \& pattern, context \& pattern), as well as the combination of all three categories (\ie code \& context \& pattern). Details can be found in Ye~\etal~\cite{ye2021automated}.

\subsubsection{Sequence-based Representation}
\label{sec:res_seq}
Sequence-based representation treats code snippets as a sequence of code tokens, and has been widely utilized in prior work to directly learn program semantics from the code sequence~\cite{yuan2022circle,tian2020evaluating,zhang2024appt,chen2019sequencer}.
In this work, we utilize a subword-level tokenizer~\cite{zhang2023pre} to split source code into multiple subwords based on token frequency.
For example, the word \textit{getUserName} can be split into three tokens, \ie \textit{get}, \textit{user}, \textit{name}.
This kind of tokenization granularity can obtain more specific and rich meanings of patched code snippets.
Besides, instead of adopting full words, smaller subwords are needed to build a vocabulary dictionary, which can reduce the vocabulary size significantly and is widely used in learning-based SE tasks, such as program repair~\cite{xia2022less,lutellier2020coconut}.


\subsubsection{Tree-based Representation}
Compared with the sequence-based representation, the tree-based representation further considers the syntactic structural information of source code~\cite{li2022dear,zhang2019novel}.
In this work, we consider abstract syntax tree (AST), which is a fundamental practice to represent the syntactic structure information of a code snippet and has been widely used in recent code representation learning studies~\cite{lin2022context, jiang2023knod}.

\subsubsection{Graph-based Representation}

Graph-based representation refers to extracting the graph information from the source code as code features~\cite{leclair2020improved,lou2021boosting}.
Existing code-related works have proposed multiple types of graph information to enhance code representation and address code-related tasks. 
In our study, we mainly focus on five types of graphs: control flow graph (CFG), control dependency graph (CDG), data dependency graph (DDG), program dependency graph (PDG), and code property graph (CPG). 
In particular, (1) CFG describes the execution flow of the program. For example, if statement \texttt{A} may occur right after statement \texttt{B}, then there must be an edge from \texttt{A} to \texttt{B} in CFG; (2) CDG describes control dependency in a program which happens when the branching result of a predicate determines the execution of the immediate statements; (3) DDG describes data dependency in a program which happens when a statement uses variables whose value depends on the execution of the previous statement; (4) PDG combines data dependency with control dependency; and (5) CPG combines several graph representations, \eg CFG, CDG, and DDG into one graph structure, thus containing more graph edge and node information.
Following prior work~\cite{siow2022learning}, we use Joern~\cite{yamaguchi2014modeling} to construct graph representations and DGL~\cite{wang2019deep} to perform graph calculation.

Overall, we consider 15 code representations, including eight heuristic-based ones (\ie TF-IDF, code, context, pattern, code \& context, code \& pattern, context \& pattern, and code \& context \& patterns), one sequence-based (\ie sub-word code sequence), one tree-based (\ie AST), and five graph-based (\ie CFG, CDG, DDG, PDG and CPG) features.

\subsection{Model Selection}
\label{sec:model_select}

Following previous work~\cite{siow2022learning,ye2021automated,tian2020evaluating}, we select models based on the corresponding types of code representations, detailed as follows.

To evaluate the effectiveness of heuristic-based representation, we select three machine learning algorithms, \ie Support Vector Machine (SVM)~\cite{cortes1995support}, Naive Bayes~\cite{hand2001idiot} and XGBoost~\cite{chen2016xgboost}. 
All of them have been adopted and shown impressive performance in previous APCA studies~\cite{ye2021automated,lin2022context,tian2020evaluating}.
In particular, Naive~Bayes first calculates the probability of which class a sample belongs to based on independent features, then selects the most probable one as its decision. 
SVM attempts to find a hyperplane that can distinctly identify samples in different classes.
Besides, XGBoost is an implementation of a gradient-boosting machine that is known for scalability and performance.

To evaluate the effectiveness of sequence-based representation, we choose Long-Short Term Memory (LSTM), Bi-directional LSTM (BiLSTM) network, and the encoder part of the Transformer \cite{vaswani2017attention}.
In particular, LSTM is a specialized recurrent neural network (RNN) for modeling long-term dependencies of sequences.
Thus, LSTM is well-suited to extract the contextual semantic features containing token sequential dependencies between buggy and patched code snippets~\cite{zhang2024appt}.
Compared with LSTM, BiLSTM takes advantage of both forward and backward context, thus modeling the sequential dependencies between tokens in both directions of the sequence.
Besides, the transformer employs multi-head attention, with the query-key mechanism to emphasize contextual information when encoding source code.
\looseness=-1

To evaluate the effectiveness of tree-based representation, we select two widely-aopted tree-aware model, \ie Tree-Structured Long Short-Term Memory Networks (TreeLSTM)~\cite{tai2015improved} and Tree-based Convolutional Neural Network (TBCNN)~\cite{mou2016convolutional}.
In particular, TreeLSTM is a variant of the basic LSTM architecture to model tree-structured topologies, as LSTM only allows for strictly sequential information propagation.
Unlike a single forget gate in the standard LSTM unit, the TreeLSTM unit contains one forget gate for each child.
Thus, the Tree-LSTM unit is able to selectively incorporate information from each child and learn to emphasize semantic heads in our patch correctness assessment task.
TBCNN~\cite{mou2016convolutional} considers that nodes with similar structures in AST should be represented in a similar way and employs continuous binary trees to model a node's representation from its children through a single neural layer.
In our work, we re-implement TBCNN with Pytorch framework and optimize the code's CPU-heavy part to make TBCNN process large-scale tree input (5000 for maximum tree nodes, 200 for maximum children nodes), so as to support our patch correctness assessment task. 


To evaluate the effectiveness of graph-based representation, we select three popular and widely-used graph neural network models, \ie Graph Convolutional Network (GCN)~\cite{kipf2016semi}, Graph Attention Network (GAT)~\cite{velickovic2017graph}, and Gated Graph Neural Network (GGNN)~\cite{li2015gated}. 
In particular, GCN captures the dependencies and features of nodes and their neighbors by applying convolutional processes to graphs.
Thus, the neighboring information of a node is aggregated using convolutional operation and layers of networks can be stacked to enhance the learning of node features.
GAT attempts to incorporate an attention mechanism in GNNs to attend over the neighborhood of a node to capture the local neighborhood information.
GGNN further aggregates node information by Gated Recurrent Unit (GRU) at every iteration to learn the node representations.

Overall,  we select eleven models for patch assessment, including three heuristic-based models (\ie NaiveBayes, XgBoost, and SVM), three sequence-based models (\ie LSTM, BiLSTM, and Transformer), two tree-based models (\ie TreeLSTM, TBCNN) and three graph-based models (\ie GCN, GAT, and GGNN).

\subsection{Evaluation Metrics}
Following previous studies~\cite{lin2022context,tian2020evaluating,le2023invalidator,zhang2024appt}, the objective of our study is to filter out overfitting patches among plausible patches.
We evaluate the performance of investigated APCA approaches by comparing the predicted labels with the ground truth labels, which can be categorized into four scenarios, \ie True Positive (TP), False Positive (FP), False Negative (FN), and True Negative (TN).
In particular, TP refers to an overfitting patch that is identified as overfitting; FP refers to a correct patch that is identified as overfitting; FN refers to an overfitting patch that is identified as correct; and TN refers to a correct patch that is identified as correct.
Then, we select five standard metrics to evaluate APCA approaches, defined as follows.

\textit{Accuracy} measures the proportion of correctly reported (whether the patch is correct or not) patches.

\begin{equation}
Accuracy = \frac{TP + TN}{TP + TN + FP + FN}
\end{equation}

\textit{Recall} measures the ratio of reported overfitting patches over all the real overfitting patches.
\begin{equation}
Recall = \frac{TP }{TP + FN}
\end{equation}

\textit{Precision} measures the proportion of real overfitting patches over the reported overfitting patches.
\begin{equation}
Precision = \frac{TP }{TP + FP}
\end{equation}

\textit{F1} measures twice the multiplication of precision and recall divided by the sum of them.

\begin{equation}
F1 = 2 * \frac{Precision * Recall }{Precision + Recall}
\end{equation}

\textit{Area Under Curve (AUC)} measures the area under the receiver operating characteristic curve.
Here, $rank\_i$ represents the predicted rank of the $i$-th positive sample, $P$ represents the number of positive samples, and $N$ represents the number of negative samples.
\begin{equation}
AUC = \frac{\sum_{i=1}^{P}rank_i - \frac{P(P+1)}{2}}{P\times N}
\label{5}
\end{equation}

\subsection{Datasets}



Following previous studies~\cite{tian2020evaluating,lin2022context}, We focus on Defects4J-v2.0~\cite{just2014defects4j} with 835 bugs from 17 open-source projects, which is the most widely-adopted benchmark in APR and APCA research~\cite{zhang2023survey}.
We collect plausible patches generated by APR tools from latest studies, including Liu~\etal~\cite{liu2020efficiency}, Xiong~\etal~\cite{xiong2018identifying}, Ali~\etal~ \cite{ghanbari2022patch}, Tian~\etal~\cite{tian2020evaluating,tian2022predicting,tian2022best}, Lin~\etal~\cite{lin2022context}. 
We also scan the artifacts released in the APR literature towards identifying more plausible patches generated by recent APR tools.
To address the imbalance risk, we include patches
from developers that are correct according to prior work~\cite{tian2020evaluating}.
We then perform a filtering process to discard duplicate patches.
In particular, we restore these patches to corresponding buggy code snippets and patched code snippets, and compare them after removing white spaces and comments from the code snippets. 
If the code snippets are exactly the same, the corresponding patches are judged as duplicates.
Table~\ref{tab:datasets} provides the statistics on the collected patches.
Overall, we share with the community the largest patch dataset for Defects4J to date, which includes 2274 (1169 overfitting and 1105 correct) plausible patches.

\begin{table}[htbp]
  \footnotesize
  \centering
  \caption{Statistics of patches in our dataset}
    \begin{tabular}{c|c|cc|c||c|c|cc|c}
    \toprule
    ID    & APR Tool & \# Correct & \# Overfitting & \# Total & ID    & APR Tool & \# Correct & \# Overfitting & \# Total \\
    \midrule
    1     & 3sFix & 1     & 65    & 66    & 21    & KaliA & 3     & 40    & 43 \\
    2     & ACS   & 32    & 13    & 45    & 22    & LSRepair & 2     & 11    & 13 \\
    3     & AVATAR & 6     & 25    & 31    & 23    & Nopol & 1     & 27    & 28 \\
    4     & Arja  & 16    & 187   & 203   & 24    & Nopol2015 & 7     & 29    & 36 \\
    5     & Arja-e & 0     & 48    & 48    & 25    & Nopol2017 & 0     & 70    & 70 \\
    6     & CapGen & 9     & 39    & 48    & 26    & PraPR & 1     & 13    & 14 \\
    7     & Cardumen & 0     & 8     & 8     & 27    & RSRepair & 0     & 8     & 8 \\
    8     & ConFix & 4     & 58    & 62    & 28    & RSRepairA & 4     & 34    & 38 \\
    9     & DeepRepair & 4     & 8     & 12    & 29    & SOFix & 2     & 1     & 3 \\
    10    & Developer & 965   & 0     & 965   & 30    & SequenceR & 8     & 46    & 54 \\
    11    & DynaMoth & 1     & 21    & 22    & 31    & SimFix & 21    & 34    & 55 \\
    12    & Elixir & 7     & 14    & 21    & 32    & SketchFix & 3     & 8     & 11 \\
    13    & FixMiner & 3     & 26    & 29    & 33    & TBar  & 14    & 50    & 64 \\
    14    & GenProg & 1     & 26    & 27    & 34    & genPat & 0     & 1     & 1 \\
    15    & GenProgA & 0     & 29    & 29    & 35    & jGenProg & 5     & 34    & 39 \\
    16    & HDRepair & 7     & 2     & 9     & 36    & jKali & 3     & 20    & 23 \\
    17    & Hercules & 0     & 4     & 4     & 37    & jMutRepair & 0     & 12    & 12 \\
    18    & JGenProg2015 & 1     & 6     & 7     & 38    & kPAR  & 3     & 30    & 33 \\
    19    & Jaid  & 32    & 38    & 70    & 39    & ssFix & 3     & 5     & 8 \\
    20    & Kali  & 0     & 15    & 15    & \multicolumn{2}{c|}{Our Dataset} & 1169    & 1105   & 2274 \\
    \bottomrule
    \end{tabular}%
  \label{tab:datasets}
\end{table}%

\subsection{Implementation Details}

We implement all approaches based on the Pytorch framework.
For model training, we consider the default hyper-parameters and model architecture following previous work \cite{siow2022learning}.
In particular, the learning rate is set to 0.001 for all models except for the transformer (0.0005) due to training convergence.
The maximum epoch is set to 100 to ensure models would be well trained. 
The batch size is 128 for most model training except for TBCNN. 
Due to TBCNN's support of the representation of large trees, either RAM or GPU's memory may run out. 
In our practice, we set the batch size to 16 for TBCNN. 
The word embedding dimension is 128, and the dropout rate is 0.2.
We randomly split the whole dataset into training, validation, and evaluation to train, tune, and test all models.
We then report the performance of all the models with a five-fold cross-validation following previous studies~\cite{lin2022context,tian2020evaluating}.
For feature extraction, we calculate the TF-IDF feature vector according to token frequency.
We use Coming~\cite{martinez2019coming} to extract code and context features, and ADD~\cite{madeiral2018towards} to extract repair features following~\cite{ye2021automated}.
We use Joern~\cite{yamaguchi2014modeling} to build tree and graph representations and DGL~\cite{wang2019deep} to implement the corresponding approaches, including graph data structure and graph calculation.
All experiments are conducted on one Ubuntu 18.04.3 server equipped with two Tesla V100-SXM2-32GB GPUs.

\section{Evaluation and Results}
\label{sec:re&an}

\subsection{RQ1: Comparison with Different Code Representations}
\label{sec:rq1}

\textbf{Experimental Design.}
In this section, we attempt to investigate the performance of different code representations in predicting patch correctness.
We consider four types of representations and their corresponding classification models, as illustrated in Sections~\ref{sec:representation} and \ref{sec:model_select}.
In particular, we conduct our experiment to systematically evaluate the following 45 APCA models, \ie (1) eight heuristic-based representations (TF-IDF, Code, Context, Pattern, Code-Context, Code-Pattern, Context-Pattern,  Code-Context-Pattern) $\times$ three classification models (Naive Bayes, SVM and XGBoost); 
(2) one sequence-based representation $\times$ three model (LSTM, BiLSTM and Transformer);
(3) one tree-based representation $\times$ two models (TreeLSTM and TBCNN);
and (4) five graph-based representations (CFG, DDG, PDG, CDG, and CPG) $\times$ three models (GAT, GCN and GGNN).

\begin{table*}[htbp]
  \centering
  \caption{Effectiveness of different heuristic-based representation features}
  \resizebox{0.9\linewidth}{!}{
    \begin{tabular}{c|c|ccccc|c}
    \toprule
    Model & Feature Type & Accuracy & Precision & Recall & F1    & AUC   & Average \\
    \midrule
    \multirow{8}[0]{*}{NaiveBayes} & TF-IDF & \textbf{79.23\%} & \textbf{76.77\%} & \textbf{82.22\%} & \textbf{79.39\%} & \textbf{79.30\%} & \textbf{79.38\%} \\
          & Code  & 65.04\% & 65.40\% & 59.93\% & 62.50\% & 64.91\% & 63.56\% \\
          & Context & 69.08\% & 69.30\% & 65.71\% & 67.41\% & 69.00\% & 68.10\% \\
          & Pattern & 65.66\% & 67.40\% & 57.49\% & 61.94\% & 65.44\% & 63.59\% \\
          & Code \& Context & 67.98\% & 67.48\% & 66.16\% & 66.78\% & 67.94\% & 67.27\% \\
          & Code \& Pattern & 67.33\% & 66.03\% & 67.78\% & 66.87\% & 67.34\% & 67.07\% \\
          & Context \& Pattern & 71.06\% & 70.10\% & 70.85\% & 70.43\% & 71.05\% & 70.70\% \\
          & Code \& Context\& Pattern & 69.34\% & 68.22\% & 69.41\% & 68.78\% & 69.35\% & 69.02\% \\
    \midrule
    \multirow{8}[1]{*}{SVM} & TF-IDF & \textbf{61.09\%} & \textbf{56.14\%} & 93.95\% & \textbf{70.20\%} & \textbf{61.95\%} & \textbf{68.67\%} \\
          & Code  & 49.01\% & 48.79\% & 97.03\% & 64.88\% & 50.27\% & 62.00\% \\
          & Context & 54.81\% & 52.04\% & 93.05\% & 66.72\% & 55.81\% & 64.49\% \\
          & Pattern & 51.34\% & 50.08\% & 94.22\% & 65.30\% & 52.46\% & 62.68\% \\
          & Code \& Context & 54.76\% & 52.86\% & 90.35\% & 65.84\% & 55.69\% & 63.90\% \\
          & Code \& Pattern & 50.02\% & 49.35\% & \textbf{98.83\%} & 65.82\% & 51.30\% & 63.06\% \\
          & Context \& Pattern & 54.46\% & 51.99\% & 91.86\% & 66.12\% & 55.44\% & 63.97\% \\
          & Code \& Context\& Pattern & 54.46\% & 51.99\% & 91.86\% & 66.12\% & 55.44\% & 63.97\% \\
    \midrule
    \multirow{8}[2]{*}{XGBoost} & TF-IDF & \textbf{80.41\%} & \textbf{78.07\%} & \textbf{83.21\%} & \textbf{80.50\%} & \textbf{80.48\%} & \textbf{80.53\%} \\
          & Code  & 70.36\% & 70.31\% & 67.69\% & 68.97\% & 70.29\% & 69.52\% \\
          & Context & 72.86\% & 71.00\% & 74.82\% & 72.84\% & 72.91\% & 72.89\% \\
          & Pattern & 67.41\% & 68.58\% & 61.19\% & 64.62\% & 67.25\% & 65.81\% \\
          & Code \& Context & 73.52\% & 72.27\% & 74.01\% & 73.11\% & 73.53\% & 73.29\% \\
          & Code \& Pattern & 68.52\% & 68.16\% & 75.26\% & 70.17\% & 68.69\% & 70.16\% \\
          & Context \& Pattern & 75.28\% & 73.64\% & 76.63\% & 75.10\% & 75.31\% & 75.19\% \\
          & Code \& Context\& Pattern & 75.85\% & 74.49\% & 76.63\% & 75.53\% & 75.87\% & 75.67\% \\
    \bottomrule
    \end{tabular}%
    }
  \label{tab:rq1_engineering}%
\end{table*}%

\textbf{Results of heuristic-based Representation.}
Table~\ref{tab:rq1_engineering} presents the comparison results of heuristic-based representation in terms of five metrics, involving eight features and three models.
When comparing different features, we find that TF-IDF consistently offers superior performance among all metrics and classifiers, except Recall on SVM, as highlighted in bold within Table~\ref{tab:rq1_engineering}.
The average performance of TF-IDF across three classifiers achieves 73.58\% for accuracy, 70.33\% for precision, 86.46\% for recall, 76.70\% for F1 and 73.91\% for AUC, improving other features by 14.61\%, 11.09\%, 12.03\%, 12.96\% and 14.52\%, respectively.
For the remaining seven code-aware features, we find that, compared with single features, the combination of these features usually achieves better performance.
For example,  Code \& Context, Code \& Pattern and Context \& Pattern achieve an average of 70.21\% for accuracy, 69.96\% for precision,	67.90\% for recall, 68.81\% for F1 and 70.15\% for AUC with the XGBoost model, improving Code, Context and Pattern by 3.18\%, 1.99\%, 10.90\%, 5.79\% and 3.36\%, respectively.
Besides, the combination of three single features (\ie Code \& Context \& Pattern) shows promising performance, particularly in the XGBoost model where all five metrics are optimal among all seven code-aware features, \eg with an accuracy of 75.85\% and an F1 score of 75.52\%.	
When comparing different classifiers, we find that XGBoost, which is utilized by ODS~\cite{ye2021automated},  demonstrates superior overall performance against NaiveBayes and SVM.
For example, XGBoost achieves an accuracy of 73.03\% and an F1 score of 72.61\% for all features on average, outperforming NaiveBayes by 5.32\% and 6.75\%, SVM by 35.88\% and 9.39\%, respectively.


\begin{table*}[htbp]
  \centering
  \caption{Effectiveness of sequence-based representation on APCA}
    \begin{tabular}{c|ccccc|c}
    \toprule
    Model & Accuracy & Precision & Recall & F1    & AUC   & Average \\
    \midrule
    LSTM  & 81.60\% & 79.04\% & 84.66\% & 81.74\% & 81.68\% & 81.74\% \\
    BiLSTM & 82.04\% & 78.48\% & 86.91\% & 82.45\% & 82.16\% & 82.41\% \\
    Transformer & \textbf{82.48\%} & \textbf{78.97\%} & \textbf{87.27\%} & \textbf{82.87\%} & \textbf{82.60\%} & \textbf{82.84\%} \\
    \bottomrule
    \end{tabular}%
  \label{tab:rq1_sequence}%
\end{table*}%

\textbf{Results of Sequence-based Representation.}
Table~\ref{tab:rq1_sequence} presents the comparison results of sequence-based representation across five metrics and three models.
We find that the Transformer model achieves the highest accuracy of 82.48\%, precision of 78.97\%, recall of 87.27\%, F1 of 82.87\%, and AUC of 82.60\%.
The Transformer's superior performance can be attributed to its advanced attention mechanism that allows it to focus on different parts of the code sequence effectively, thus making it particularly suited for handling complex patterns in patched code snippets.
BiLSTM achieves impressive performance in terms of accuracy, precision, Recall, F1, and AUC, with respective scores of 82.04\%, 78.48\%, 86.91\%, 82.45\% and 82.16\%, four of which are better than those of LSTM. 
The improvement indicates the BiLSTM's enhanced capability in capturing the temporal dependencies in both forward and backward directions, which is beneficial for predicting patch correctness.
Overall, the Transformer model emerged as the most effective in sequence-based representation for APCA, achieving a well-balanced performance across all evaluated metrics. 

\begin{table*}[htbp]
  \centering
  \caption{Effectiveness of tree-based representation on APCA}
    \begin{tabular}{c|ccccc|c}
    \toprule
    Model & Accuracy & Precision & Recall & F1    & AUC   & Average \\
    \midrule
    TreeLSTM & \textbf{82.94\%} & \textbf{80.01\%} & \textbf{86.73\%} & \textbf{83.18\%} & \textbf{83.03\%} & \textbf{83.18\%} \\
    TBCNN & 81.72\% & 79.24\% & 84.62\% & 81.84\% & 81.79\% & 81.84\% \\
    \bottomrule
    \end{tabular}%
  \label{tab:rq1_tree}%
\end{table*}%

\textbf{Results of Tree-based Representation.}
Table~\ref{tab:rq1_tree} presents the comparison results of tree-based representation across five metrics and two models.
We find that TreeLSTM outperforms TBCNN across all evaluated metrics.
Specifically, TreeLSTM achieves 82.94\% accuracy, 80.01\% precision, 86.73\% recall, 83.18\% F1, and 83.03\% AUC.
The improvement over TBCMM is 1.49\%, 0.97\%, 2.49\%, 1.64\%, and 1.52\%, indicating TreeLSTM's ability to identify more overfitting patches with a lower false-positive rate.

\begin{table*}[htbp]
  \centering
  \caption{Effectiveness of graph-based representation on APCA}
    \begin{tabular}{c|c|c|c|c|c|c|c}
    \toprule
    Graph & Model & Accuracy & Precision & Recall & F1    & AUC   & Average \\
    \midrule
    \multirow{4}[4]{*}{CFG} & GAT   & 80.86\% & 78.75\% & 85.02\% & 81.75\% & 80.83\% & 81.44\% \\
          & GCN   & 75.59\% & 76.25\% & 74.88\% & 75.47\% & 75.60\% & 75.56\% \\
          & GGNN  & 83.15\% & 82.09\% & 85.22\% & 83.60\% & 83.13\% & 83.44\% \\
\cmidrule{2-8}          & {Average} & {79.87\%} & {79.03\%} & {81.71\%} & {80.27\%} & \multicolumn{1}{c}{{79.85\%}} & {80.15\%} \\
    \midrule
    \multirow{4}[2]{*}{DDG} & GAT   & 82.37\% & 80.31\% & 84.66\% & 82.39\% & 82.42\% & 82.43\% \\
          & GCN   & 75.99\% & 75.03\% & 76.00\% & 75.46\% & 75.99\% & 75.69\% \\
          & GGNN  & 83.95\% & 81.51\% & 86.73\% & 84.04\% & 84.02\% & 84.05\% \\
          & {Average} & {80.77\%} & {78.95\%} & {82.46\%} & {80.63\%} & {80.81\%} & {80.72\%} \\
    \midrule
    \multirow{4}[4]{*}{PDG} & GAT   & 81.60\% & 79.51\% & 83.80\% & 81.56\% & 81.65\% & 81.62\% \\
          & GCN   & 75.92\% & 75.14\% & 75.74\% & 75.32\% & 75.92\% & 75.61\% \\
          & GGNN  & 82.72\% & 80.41\% & 85.19\% & 82.69\% & 82.79\% & 82.76\% \\
\cmidrule{2-8}          & {Average} & {80.08\%} & {78.35\%} & {81.58\%} & {79.86\%} & {80.12\%} & {80.00\%} \\
    \midrule
    \multirow{4}[4]{*}{CDG} & GAT   & 75.52\% & 75.45\% & 78.58\% & 76.69\% & 75.41\% & 76.33\% \\
          & GCN   & 72.37\% & 72.84\% & 74.24\% & 73.47\% & 72.30\% & 73.04\% \\
          & GGNN  & 81.67\% & 80.26\% & 85.62\% & 82.81\% & 81.53\% & 82.38\% \\
\cmidrule{2-8}          & {Average} & {76.52\%} & {76.18\%} & {79.48\%} & {77.66\%} & {76.41\%} & {77.25\%} \\
    \midrule
    \multirow{4}[4]{*}{CPG} & GAT   & 82.55\% & 79.46\% & 86.64\% & 82.87\% & 82.64\% & 82.83\% \\
          & GCN   & 81.80\% & 77.35\% & 88.63\% & 82.60\% & 81.96\% & 82.47\% \\
          & GGNN  & 83.73\% & 81.00\% & 87.09\% & 83.90\% & 83.81\% & 83.91\% \\
\cmidrule{2-8}          & {Average} & {\textbf{82.69\%}} & {\textbf{79.27\%}} & {\textbf{87.45\%}} & {\textbf{83.12\%}} & {\textbf{82.80\%}} & {\textbf{83.07\%}} \\
    \bottomrule
    \end{tabular}%
  \label{tab:rq1_graph}%
\end{table*}%

\textbf{Results of Graph-based Representation.}
Table~\ref{tab:rq1_graph} presents the comparison results of graph-based representation across five graph-based representations, three models, and five metrics.
When comparing different graphs, we find CPG achieves optimal performance, with accuracy, precision, recall, F1, and AUC being 82.69\%, 79.27\%, 87.45\%, 83.12\%, 82.80\%, and 83.07\% on average across three classifiers.
The overall improvement of CPG against CFG, DDG, PDG, and CDG across five metrics reaches 3.65\%, 2.90\%, 3.84\%, and 7.53\%, respectively.
Such improvement of CPG is reasonable, as CPG is a combination of several graphs, thereby containing more information, such as data dependency and control dependency.
The second-best performance is exhibited by CFG, DDG, and PDG, which attain comparable performance, \eg 80.15\%, 80.72\%, and 80.00\% for five metrics on average.
Besides, CDG performs the worst among the five graphs with an accuracy of only 76.56\% on average, which is 2.90\%, 3.47\%, 2.75\%, and 5.82\% less than CFG, DDG, PDG, and CPG.
When comparing the three models, we find that GGNN yields consistently optimal performance and GCN achieves the second-best performance across all graph types and metrics.
The improvement of GGNN against GCN reaches 10.43\%, 11.04\%, 9.46\%, 12.78\%, and 1.74\%, respectively.
The possible reason lies in the fact that GGNN utilizes GRU to update node representations in the graph, which allows for the incorporation of information from a node's neighbors.
Similarly, GCN outperforms GAT by 2.45\%, 1.97\%, 1.39\%, 7.92\% and 1.30\% on the five metrics.

\textbf{Comparison of Four Types of Representation.}
From Tables~\ref{tab:rq1_engineering} to~\ref{tab:rq1_graph}, we first find that sequence-based approaches have a better performance compared with the heuristic-based approaches. with an improvement of 2.57\% for accuracy and 4.88\% for recall.
Second, the difference between the sequence-based and tree-based approaches is not very obvious, \eg only 0.41\% improvement.
Third, we find that the graph-based approach achieves the best performance among representations, with an average accuracy of 79.99\%, a precision of 78.36\%, a recall of 82.54\%, an F1 of 80.31\%, and an AUC of 80.00\% across all graphs and classifiers.
For example, GGNN with CGP outperforms XFBoost with TF-IDF, transformer, and TreeLSTM by 4.19\%, 1.29\%, and 0.88\% on average, indicating that the semantic and syntactic information on graphs is beneficial in learning the semantics of the buggy program.

\begin{figure}[t]
\centering
    \subfigure[Overfitting patches identification] {
        \includegraphics[width=0.4\columnwidth]{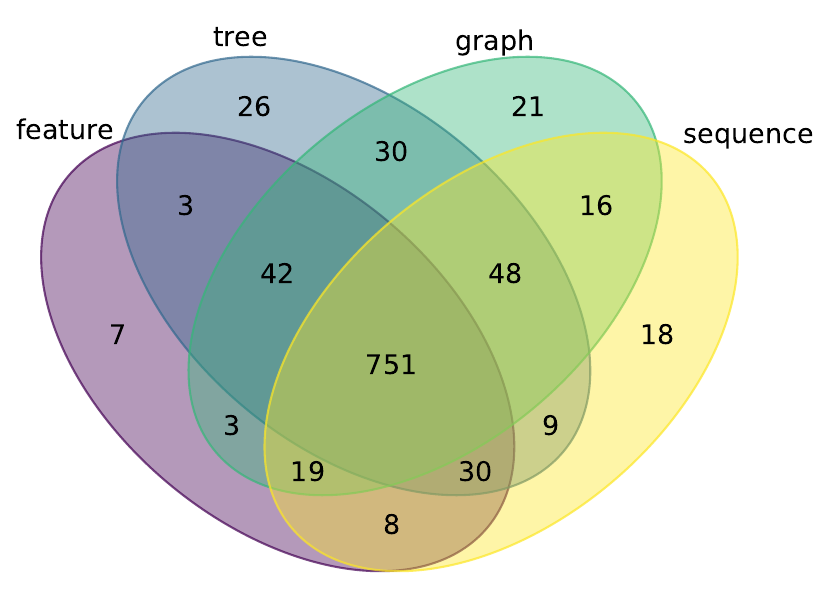}
        \label{fig:veen_old}
    }
    \subfigure[Correct patches identification] {
        \includegraphics[width=0.4\columnwidth]{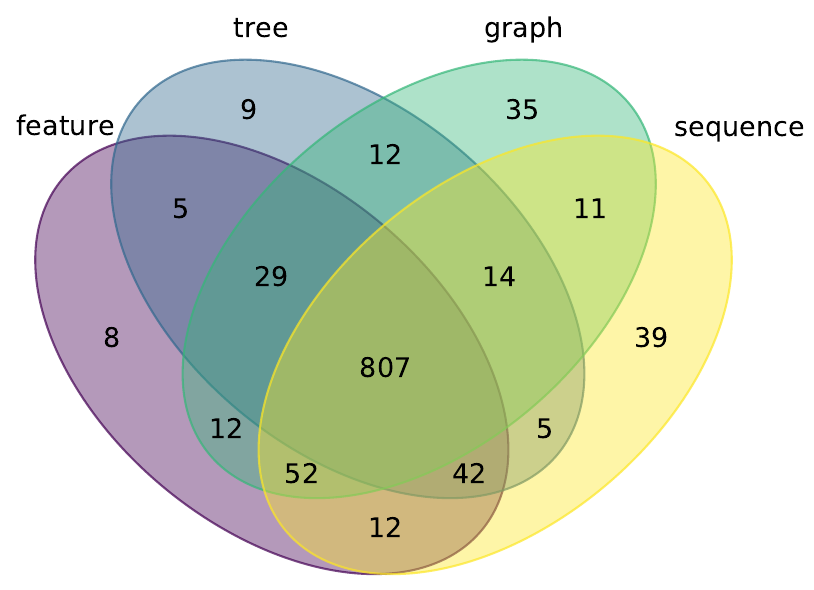}
        \label{fig:veen_new}
    }
\caption{The overlaps of unique patches correctly predicted by four representations}
\label{fig:veen}
\end{figure}

\myfinding{1}{Our results reveal that,
(1) among four types of representation, the graph-based representation achieves optimal performance with average accuracy and recall of 79.99\% and 82.54\% in predicting patch correctness;
(2) TF-IDF is superior to other heuristic-based representations, \eg TF-IDF with XGBoost has the highest accuracy of 80.41\% and 78.07\% for precision;
(3) CPG performs better than other graph-based representations by 3.65\%, 2.90\%, 3.84\%, and 7.53\% on average.
}

\subsection{RQ2: Comparison with State-of-the-art Approaches}
\label{sec:baseline}

\begin{table*}[t]
  \centering
  \caption{Comparison results of four types of representations with state-of-the-art.}
    \begin{tabular}{c|cccc|c}
    \toprule
    Approach & Accuracy & Precision & Recall & F1    & Average \\
    \midrule
    ODS   & 71.00\% & 72.00\% & 71.00\% & 71.00\% & 71.25\% \\
    Tian~\etal (LR)    & 75.00\% & 77.78\% & 74.00\% & 75.00\% & 75.45\% \\
    Tian~\etal (DT)    & 63.74\% & 65.09\% & 65.37\% & 65.23\% & 64.86\% \\
    Tian~\etal (SVM)   & 76.58\% & 78.48\% & 75.76\% & 77.09\% & 76.98\% \\
    Tian~\etal (NB)    & 61.00\% & 64.00\% & 55.00\% & 59.00\% & 59.75\% \\
    CACHE & 74.00\% & 74.00\% & 87.00\% & 79.00\% & 78.50\% \\
    \midrule 
    Heuristic-based  & 80.41\% & 78.07\% & 83.21\% & 80.50\% & 80.55\% \\
    Sequence-based  & 82.48\% & 78.97\% & \textbf{87.27\%} & 82.87\% & 82.90\% \\
    Tree-based   & 82.94\% & 80.01\% & 86.73\% & 83.18\% & 83.03\% \\
    Graph-based  & \textbf{83.73\%} & \textbf{81.00\%} & 87.09\% & \textbf{83.90\%} & \textbf{83.81\%} \\
    \midrule
    PATCH-SIM & \textbf{66.00\%} & 100.00\% & \textbf{57.00\%} & \textbf{73.00\%} & \textbf{74.00\%} \\
    Tree  & 80.00\% & \textbf{84.00\%} & 91.00\% & 88.00\% & 85.75\% \\
    \bottomrule
    \end{tabular}%
  \label{tab:rq2_baseline}%
\end{table*}%

\textbf{Experimental Design.}
In this section, we attempt to compare the four types of representations against previous state-of-the-art APCA approaches.
We consider the best-performing representation from each of the four types as representative approaches, \ie TF-IDF with XGBoost, the sequence with transformer, AST with TreeLSTM, and CPG with GGNN.
We then select four state-of-the-art APCA approaches as baselines, summarized as follows.

\begin{itemize}
    \item PATCH-SIM~\cite{xiong2018identifying}.
    Xiong~\etal~\cite{xiong2018identifying} filter overfitting patches based on the dynamic traces of test cases.
    PATCH-SIM adopts Randoop to generate extra test cases to collect execution information by default.

    \item ODS~\cite{ye2021automated}.
    Ye~\etal~\cite{ye2021automated} first extract 202 hand-crafted features and then utilize supervised learning to train a binary classification model.
    ODS serves as the pioneering effort to explore the potential of heuristic-based features for patch correctness assessment.
    
    \item Tian~\etal~\cite{tian2020evaluating}.
    It utilizes representation learning models to produce embedding for buggy and patched code snippets and then train supervised learning classifies to predict patch correctness.
    Following previous work~\cite{tian2020evaluating,tian2022best,tian2022predicting}, we select BERT as the embedding model 
    and Logistic Regression (LR), Decision Tree (DT), Naive Bayes (NB), and Support Vector Machine (SVM) as classification models.
    
    \item CACHE~\cite{lin2022context}.
    Lin~\etal~\cite{lin2022context} utilize context-aware code change embedding in reasoning about patch correctness.
    CACHE considers both program structures and code context by parsing the patched code snippet into AST representation and then utilizing the AST path technique to capture the structure information

\end{itemize}


\textbf{Results.}
Table~\ref{tab:rq2_baseline} presents the comparison results of four types of representation and four state-of-the-art APCA approaches.
We report the performance of baselines and four types by executing them on our datasets with a five-fold cross-validation.
The detailed results are illustrated in 2-6 rows (\ie ODS, LR, DT, SVM, NB and CACHE) and 7-10 rows (\ie TF-IDF+XGBoost, Transformer+sequence, AST+TreeLSTM, DDG+GGNN) in Table~\ref{tab:rq2_baseline}.
However, it is time-consuming to generate and execute additional test cases for PATCH-SIM, such as approximately 70 hours to classify 139 patches~\cite{xiong2018identifying}.
Thus, following previous studies~\cite{ye2021automated,tian2020evaluating,le2023invalidator}, we evaluate four types of representation on the 139 testing patches provided by Xiong~\etal~\cite{xiong2018identifying} and utilize the remaining patches as the training set.
We only train a tree-based model due to page limit and huge training time.
The results of the tree-based model and PATCH-SIM are illustrated in the last two rows in Table~\ref{tab:rq2_baseline}.
It is noteworthy that we include four metrics to maintain consistency with the reported results of PATCH-SIM.

First, when comparing with ODS, we find four types of representation continuously achieve better performance, with an average improvement of 13.05\%, 16.35\%, 16.53\%, and 17.63\%.
Importantly, it can be observed that, as a representative in heuristic-based representation, TF-IDF outperforms ODS by 13.25\% for accuracy, 8.43\% for precision, 17.20\% for recall, and 13.38\% for F1, indicating that TF-IDF, as a widely-used feature in many SE tasks~\cite{siow2022learning}, remains effective in predicting patch correctness. 
Second, when comparing with Tian~\etal~\cite{tian2020evaluating}, we find SVM achieves optimal performance among four variants, improving LR, DT, and NB by 2.03\%, 18.69\%, and 8.04\%, respectively.
However, four types of representations achieve better performance than SVM by 4.64\%, 7.69\%, 7.86\%, and 8.88\%, and even than the context-aware approach CACHE by 2.61\%, 5.60\%, 5.77\%, and 6.76\%.
Third, when comparing with PATCH-SIM, we find that tree-based representation achieves 80.00\% for accuracy, 84.00\% for precision, 91.00\% for recall, and 88.00\% for F1, three of which are superior, except precision.
We think the results about precision are acceptable as dynamic execution information naturally enhances the precision, while representation-based approaches only rely on static information.

\myfinding{2}{Our comparison results demonstrate that,
(1) the heuristic-based approach outperforms ODS by 8.43\% $\sim$ 17.20\% across four metrics;
(2) the sequence-based approach outperforms four BERT-like approaches, \eg with an improvement of 4.64\% over SVM;
(3) the tree-based and graph-based approaches outperform CACHE and PATCH-SIM on all metrics, except precision for PATCH-SIM.
}

\subsection{RQ3: Effectiveness of Fusion-based APCA Approach}
\label{sec:fusion}

\textbf{Experimental Design.}
As mentioned in Section~\ref{sec:rq1}, different representations are able to complement each other in identifying overfitting patches.
Different representations capture various aspects of source code, such as semantic and structural information.
Thus, in this section, we investigate the potential of combining these features to improve the prediction performance.
We consider two strategies to combine different representations according to the stage of fusion, \ie intermediate fusion and backend fusion.
The first one attempts to concatenate feature vectors to train prediction models, and the second one performs ensemble learning on already trained prediction models.

To perform the intermediate fusion, we select the best-performing representation and its corresponding model from heuristic-based, sequence-based, tree-based, and graph-based representations and fuse different representations by vector concatenation.
We then train a binary classifier to predict the patch correctness based on the concatenated vectors.
Following previous studies~\cite{zhang2024appt,lin2022context}, we apply a standard softmax function to obtain the probability distribution over correctness.
To perform the backend fusion, we apply the stacking method in ensemble learning to the patch correctness assessment. 
We select four best-performing already-trained models from four types of representation and record the softmax function outputs during inference.
We then compute the average probability of these models as the final prediction result.
As a result, we extend our experiments to evaluate 22 APCA models, including 11 combined features $\times$ 2 fusion strategies.

\begin{table*}[t]
  \centering
  \caption{Comparison results of representation fusion}
  \resizebox{0.99\linewidth}{!}{
    \begin{tabular}{c|ccccc|c|ccccc|c}
    \toprule
    \multicolumn{1}{c|}{\multirow{2}[4]{*}{Representation}} & \multicolumn{6}{c|}{Intermediate Fusion}      & \multicolumn{6}{c}{Backend Fusion} \\
\cmidrule{2-13}          & Accuracy & Precision & Recall & F1    & AUC   & Average   & Accuracy & Precision & Recall & F1    & AUC   & Average \\
    \midrule
    Heu-Seq & \textbf{84.20\%} & \textbf{83.25\%} & 87.05\% & \textbf{85.06\%} & \textbf{84.10\%} & \textbf{84.73\%} & \textbf{85.42\%} & \textbf{84.00\%} & 88.75\% & \textbf{86.29\%} & \textbf{85.30\%} & \textbf{85.95\%} \\
    Heu-Tre & 83.04\% & 80.96\% & 88.11\% & 84.31\% & 82.85\% & 83.85\% & 81.04\% & 78.14\% & 88.66\% & 82.94\% & 80.76\% & 82.31\% \\
    Heu-Gra & 84.04\% & 83.04\% & 87.05\% & 84.94\% & 83.93\% & 84.60\% & 82.76\% & 81.37\% & 86.62\% & 83.89\% & 82.62\% & 83.45\% \\
    Seq-Tre & 81.71\% & 79.86\% & 86.51\% & 83.03\% & 81.53\% & 82.53\% & 80.27\% & 78.37\% & 85.76\% & 81.79\% & 80.07\% & 81.25\% \\
    Seq-Gra & 82.87\% & 81.88\% & 86.09\% & 83.83\% & 82.76\% & 83.49\% & 81.93\% & 79.75\% & 87.26\% & 83.31\% & 81.74\% & 82.80\% \\
    Tre-Gra & 83.87\% & 81.56\% & \textbf{89.07\%} & 85.12\% & 83.68\% & 84.66\% & 83.26\% & 81.91\% & \textbf{86.83\%} & 84.29\% & 83.13\% & 83.88\% \\
    \midrule
    Heu-Seq-Tre & 82.10\% & 81.20\% & 85.23\% & 83.11\% & 81.98\% & 82.72\% & 82.21\% & 80.24\% & 87.36\% & 83.58\% & 82.02\% & 83.08\% \\
    Heu-Seq-Gra & 83.48\% & 81.40\% & 88.33\% & 84.69\% & 83.30\% & 84.24\% & 83.26\% & 82.10\% & 86.51\% & 84.24\% & 83.14\% & 83.85\% \\
    Heu-Tre-Gra & 83.54\% & 81.03\% & 89.08\% & 84.85\% & 83.33\% & 84.37\% & 83.54\% & 82.27\% & 87.04\% & 84.56\% & 83.41\% & 84.16\% \\
    Seq-Tre-Gra & 82.48\% & 80.50\% & 87.48\% & 83.76\% & 82.30\% & 83.30\% & 82.37\% & 80.70\% & 86.83\% & 83.58\% & 82.21\% & 83.14\% \\
    Heu-Seq-Tre-Gra & 83.26\% & 82.38\% & 86.08\% & 84.18\% & 83.16\% & 83.81\% & 83.81\% & 81.85\% & 88.33\% & 84.95\% & 83.65\% & 84.52\% \\
    \bottomrule
    \end{tabular}
    }
  \label{tab:fusion}%
\end{table*}%

\textbf{Results.}
Table~\ref{tab:fusion} illustrates the comparison results of our fusion strategies.
The first column lists 11 types of representation combinations.
The second to six columns list the results of five metrics for intermediate fusion, and the remaining columns list the corresponding results of backend fusion.
Eng, Seq, Tre, and Gra represent heuristic-based, sequence-based, tree-based, and graph-based representations.
Overall, the heuristic-sequence representation achieves an average performance of 84.73\% and 85.95\% across five metrics, which are optimal among 11 types of representation combinations for both fusion methods.

First, we find that fusion can further boost heuristic-based and sequence-based representations.
For example, for the intermediate fusion, by combining sequence-based, tree-based, and graph-based representations, the heuristic-based representation is improved by 11.97\%, 10.81\%, and 11.80\% on five metrics, respectively.
Similarly, for the backend fusion,  the improvement of heuristic-sequence, heuristic-tree, and heuristic-graph representations against the single heuristic-based representation achieves 13.58\%, 8.77\%, and 10.28\% on average.
However, for tree-based and graph-based representations, fusion fails to always bring better results. 
For example, the accuracy of the tree-based representation is improved by 1.62\% with the integration of the heuristic-based representation, while being decreased by 0.01\% with the integration of the heuristic-based representation.
The possible reason may be that tree-based and graph-based representations have already achieved impressive performance before fusion, which is difficult to be further improved compared with two relatively poor-performing representations, \ie~heuristic-based and sequence-based representations.

Second, we find that combining more than two representations does not yield further improvements.
For example, heuristic-sequence-tree is able to achieve a prediction accuracy of 82.10\%, outperforming sequence-tree by 0.48\%, while underperforming heuristic-sequence and heuristic-tree by 2.49\% and 1.13\%.
The possible reason is that the fusion methods (\ie the vector concatenation in intermediate fusion and the average prediction probability in backend fusion) employed are naive to extract the deep relationships among the various representations.
Thus, future researchers should further explore advanced techniques that can fuse various patch representations for better prediction performance.
Third, when comparing the two fusion strategies, we find that the intermediate fusion outperforms the backend fusion by 1.22\% of accuracy, 0.75\% of precision, 1.70\% of recall, 1.23\% of F1, and 1.20\% of AUC on the heuristic-sequence representation.
However, for the other two representation fusion scenarios, the backend fusion achieves better or slightly better results than the intermediate fusion, with average improvements of 1.55\%, 1.15\%, 1.28\%, 0.69\%, and 0.78\% on the heuristic-tree, heuristic-graph, sequence-tree, sequence-graph, and tree-graph representations.
Besides, for the fusion of more than two features, the two fusion methods achieve comparable results, such as 0.36\%, 0.39\%, 0.20\%, 0.17\%, 0.71\% improvement or disimprovement on five representations, as listed in the last five rows in Table~\ref{tab:fusion}.

\myfinding{3}{Our comparison results demonstrate that the fusion of two representations can improve the prediction performance, while incorporating more representations tends to have a detrimental effect due to the simplistic nature of the utilized fusion models.
}

\section{Discussion}
\label{sec:dis}

\subsection{Impact of Node Embedding}
\label{sec:node}

\textbf{Objective.}
As discussed in Section~\ref{sec:rq1}, tree-based and graph-based representations have demonstrated impressive performance compared with other representations and state-of-the-art APCA approaches.
The tree-based and graph-based representations consist of nodes representing code elements and edges representing relationships between nodes.
In particular, each node in the tree and graph is composed of two parts, the node type and the node textual information.
The former attempts to distinguish the node from the others, and the latter is a small fragment of code from the original program (the first line in each node). 
In our experiment, we utilize both the node type and the textual information to obtain an embedding that represents each node by default.
However, it is unclear whether they both have a positive effect on the prediction results.
 
\textbf{Design.}
In this section, we study the impact of different information that is embedded into the node for both tree and graph representations.
To get node representation for model learning, we consider three node embedding methods: type embedding, textual embedding, and hybrid embedding.
In type embedding, we embed each node by its type information. Nodes with different types are fed into a linear layer to learn the corresponding node representation.
In textual embedding, we first tokenize the small snippet of code that the node contains, which is input into a BiLSTM model to learn the intermediate representation of the node based on the last hidden state of the BiLSTM.
Hybrid embedding employs a linear layer to learn the concatenated representation of textual embedding and type embedding. We put the single-length vector representing type information before the vector representing textual information to form a new vector, and then input it into the linear layer.
Finally, after these embeddings are generated, we employ the models (\eg TreeLSTM and GGNN) to learn to assess patch correctness.
As a result, we extend our experiments to evaluate 51 APCA models, including one AST $\times$ two modesl $\times$ three node, and five graphs $\times$ three modesl $\times$ three node,



\begin{table*}[htbp]
\renewcommand\arraystretch{0.6}
\footnotesize
  \centering
  \caption{Comparison results of node embedding information}
  \resizebox{0.9\linewidth}{!}{
    \begin{tabular}{c|c|c|ccccc|c}
    \toprule
    Representation & Model & Embedding & Accuracy & Recall & Precision & F1    & AUC   & Average \\
    \midrule
     \multirow{6}[3]{*}{AST} & \multirow{3}[2]{*}{TreeLSTM} & Textual & \textbf{83.38\%} & \textbf{80.44\%} & \textbf{87.18\%} & \textbf{83.63\%} & \textbf{83.47\%} & \textbf{83.62\%} \\
           &       & Type  & 82.32\% & 79.63\% & 85.65\% & 82.51\% & 82.40\% & 82.50\% \\
           &       & Hybrid & 82.94\% & 80.01\% & 86.73\% & 83.18\% & 83.03\% & 83.18\% \\
 \cmidrule{2-9}          & \multirow{3}[1]{*}{TBCNN} & Textual & \textbf{82.60\%} & 79.10\% & 87.33\% & \textbf{83.01\%} & \textbf{82.72\%} & \textbf{82.95\%} \\
           &       & Type  & 79.96\% & 78.76\% & 80.54\% & 79.64\% & 79.97\% & 79.77\% \\
           &       & Hybrid & 81.72\% & \textbf{79.24\%} & \textbf{84.62\%} & 81.84\% & 81.79\% & 81.84\% \\
 \midrule
    \multirow{9}[4]{*}{CFG} & \multirow{3}[1]{*}{GAT} & Textual & \textbf{80.86\%} & 78.75\% & \textbf{85.02\%} & \textbf{81.75\%} & \textbf{80.83\%} & \textbf{81.44\%} \\
          &       & Type  & 63.32\% & 64.65\% & 64.14\% & 63.19\% & 63.32\% & 63.72\% \\
          &       & Hybrid & 80.42\% & \textbf{78.93\%} & 83.74\% & 81.19\% & 80.39\% & 80.93\% \\
\cmidrule{2-9}          & \multirow{3}[2]{*}{GCN} & Textual & 75.05\% & \textbf{76.85\%} & 72.31\% & 74.45\% & 75.08\% & 74.75\% \\
          &       & Type  & 66.50\% & 66.93\% & 66.50\% & 66.64\% & 66.50\% & 66.61\% \\
          &       & Hybrid & \textbf{75.59\%} & 76.25\% & \textbf{74.88\%} & \textbf{75.47\%} & \textbf{75.60\%} & \textbf{75.56\%} \\
\cmidrule{2-9}          & \multirow{3}[1]{*}{GGNN} & Textual & \textbf{83.15\%} & \textbf{82.09\%} & \textbf{85.22\%} & \textbf{83.60\%} & \textbf{83.13\%} & \textbf{83.44\%} \\
          &       & Type  & 79.22\% & 78.26\% & 81.58\% & 79.81\% & 79.20\% & 79.61\% \\
          &       & Hybrid & 82.45\% & 81.43\% & 84.63\% & 82.94\% & 82.43\% & 82.78\% \\
\midrule
    \multirow{9}[4]{*}{CDG} & \multirow{3}[1]{*}{GAT} & Textual & 75.47\% & 73.09\% & \textbf{83.96\%} & \textbf{77.86\%} & 75.17\% & \textbf{77.11\%} \\
          &       & Type  & 74.08\% & 70.10\% & 86.97\% & 77.62\% & 73.63\% & 76.48\% \\
          &       & Hybrid & \textbf{75.52\%} & \textbf{75.45\%} & 78.58\% & 76.69\% & \textbf{75.41\%} & 76.33\% \\
\cmidrule{2-9}          & \multirow{3}[2]{*}{GCN} & Textual & \textbf{72.37\%} & \textbf{72.84\%} & 74.24\% & \textbf{73.47\%} & \textbf{72.30\%} & \textbf{73.04\%} \\
          &       & Type  & 70.65\% & 71.77\% & 71.14\% & 71.43\% & 70.64\% & 71.13\% \\
          &       & Hybrid & 71.89\% & 71.75\% & \textbf{75.38\%} & 73.43\% & 71.76\% & 72.84\% \\
\cmidrule{2-9}          & \multirow{3}[1]{*}{GGNN} & Textual & 80.81\% & 79.36\% & 85.00\% & 82.04\% & 80.67\% & 81.58\% \\
          &       & Type  & 80.44\% & 78.04\% & \textbf{86.55\%} & 82.07\% & 80.23\% & 81.47\% \\
          &       & Hybrid & \textbf{81.67\%} & \textbf{80.26\%} & 85.62\% & \textbf{82.81\%} & \textbf{81.53\%} & \textbf{82.38\%} \\
\midrule
    \multirow{9}[4]{*}{DDG} & \multirow{3}[1]{*}{GAT} & Textual & 82.06\% & 79.03\% & 86.01\% & 82.36\% & 82.16\% & 82.32\% \\
          &       & Type  & 76.91\% & 71.66\% & \textbf{87.18\%} & 78.60\% & 77.17\% & 78.30\% \\
          &       & Hybrid & \textbf{82.37\%} & \textbf{80.31\%} & 84.66\% & \textbf{82.39\%} & \textbf{82.42\%} & \textbf{82.43\%} \\
\cmidrule{2-9}          & \multirow{3}[2]{*}{GCN} & Textual & \textbf{75.99\%} & 75.03\% & \textbf{76.00\%} & \textbf{75.46\%} & \textbf{75.99\%} & \textbf{75.69\%} \\
          &       & Type  & 74.63\% & \textbf{73.62\%} & 74.73\% & 74.15\% & 74.63\% & 74.35\% \\
          &       & Hybrid & 75.33\% & 74.74\% & 74.64\% & 74.63\% & 75.32\% & 74.93\% \\
\cmidrule{2-9}          & \multirow{3}[1]{*}{GGNN} & Textual & 82.85\% & 79.86\% & 86.64\% & 83.11\% & \textbf{82.94\%} & 83.08\% \\
          &       & Type  & 82.68\% & \textbf{82.31\%} & 82.22\% & 82.20\% & 82.66\% & 82.41\% \\
          &       & Hybrid & \textbf{83.95\%} & 81.51\% & \textbf{86.73\%} & \textbf{84.04\%} & 84.02\% & \textbf{84.05\%} \\
\midrule
    \multirow{9}[4]{*}{PDG} & \multirow{3}[1]{*}{GAT} & Textual & 81.14\% & 78.35\% & \textbf{84.63\%} & 81.33\% & 81.24\% & 81.34\% \\
          &       & Type  & 72.05\% & 70.11\% & 74.07\% & 72.02\% & 72.11\% & 72.07\% \\
          &       & Hybrid & \textbf{81.60\%} & \textbf{79.51\%} & 83.80\% & \textbf{81.56\%} & \textbf{81.65\%} & \textbf{81.62\%} \\
\cmidrule{2-9}          & \multirow{3}[2]{*}{GCN} & Textual & \textbf{75.92\%} & 75.14\% & \textbf{75.74\%} & \textbf{75.32\%} & \textbf{75.92\%} & \textbf{75.61\%} \\
          &       & Type  & 72.77\% & 72.96\% & 70.09\% & 71.43\% & 72.70\% & 71.99\% \\
          &       & Hybrid & 75.52\% & \textbf{76.39\%} & 71.85\% & 74.03\% & 75.42\% & 74.64\% \\
\cmidrule{2-9}          & \multirow{3}[1]{*}{GGNN} & Textual & \textbf{82.72\%} & \textbf{80.41\%} & \textbf{85.19\%} & \textbf{82.69\%} & \textbf{82.79\%} & \textbf{82.76\%} \\
          &       & Type  & 80.06\% & 77.42\% & 83.33\% & 80.25\% & 80.15\% & 80.24\% \\
          &       & Hybrid & 82.45\% & 80.32\% & 84.63\% & 82.38\% & 82.51\% & 82.46\% \\
\midrule
    \multirow{9}[5]{*}{CPG} & \multirow{3}[1]{*}{GAT} & Textual & \textbf{82.55\%} & \textbf{79.46\%} & \textbf{86.64\%} & \textbf{82.87\%} & \textbf{82.64\%} & \textbf{82.83\%} \\
          &       & Type  & 78.11\% & 74.34\% & 84.84\% & 79.13\% & 78.28\% & 78.94\% \\
          &       & Hybrid & 82.06\% & 79.33\% & 85.56\% & 82.27\% & 82.15\% & 82.27\% \\
\cmidrule{2-9}          & \multirow{3}[2]{*}{GCN} & Textual & 81.27\% & \textbf{77.90\%} & 86.10\% & 81.75\% & 81.39\% & 81.68\% \\
          &       & Type  & 81.01\% & 77.62\% & 85.83\% & 81.49\% & 81.12\% & 81.41\% \\
          &       & Hybrid & \textbf{81.80\%} & 77.35\% & \textbf{88.63\%} & \textbf{82.60\%} & \textbf{81.96\%} & \textbf{82.47\%} \\
\cmidrule{2-9}          & \multirow{3}[2]{*}{GGNN} & Textual & 82.99\% & 80.90\% & 85.20\% & 82.94\% & 83.04\% & 83.01\% \\
          &       & Type  & 82.28\% & 79.55\% & 85.74\% & 82.50\% & 82.36\% & 82.49\% \\
          &       & Hybrid & \textbf{83.73\%} & \textbf{81.00\%} & \textbf{87.09\%} & \textbf{83.90\%} & \textbf{83.81\%} & \textbf{83.91\%} \\
    \bottomrule
    \end{tabular}%
    }
  \label{tab:node}%
\end{table*}%

\textbf{Results.}
Table~\ref{tab:node} presents the results of three embedding methods (\ie type, textual and hybrid embedding) for six tree-based or graph representations and five models across five metrics.
First, we find that the textual node embedding consistently has a better prediction performance than the type node embedding.
For example, GAT model with PDG achieves 81.14\% for accuracy, 78.35\% for recall, 84.63\% for precision, 81.33\% for F1, 81.24\% for AUC under the textual embedding, improving those of the type embedding by 12.62\%, 11.75\%, 14.26\%, 12.93\%, and 12.66\%, respectively.
More importantly, among 85 evaluation scenarios (calculated one AST representation * two tree models * five metrics, plus five graphs * three graph models  * five metrics), the textual embedding outperforms the type embeddings in 79 of these scenarios, with an average improvement of 5.24\%.
We think that the possible reason for the advance of the node's textual description lies in the semantic information.
The textual information of nodes often conveys the basic semantic information of the code unit, which can aid the model in better understanding the program's behavior.
For example, developers may name one function as \texttt{SetHeightValue} to indicate that this function can set the value of height as they want. 
If this name is represented as its type (\eg \texttt{FunctionDef} in AST), the critical semantic information would be missed, resulting in suboptimal model training.
Second, we find that combining the node textual information with its type (\ie the hybrid embedding) can further increase the patch correctness prediction performance.
For example, when the hybrid embedding is applied to CPG, GGNN achieves 83.73\% for accuracy,	81.00\% for precision, 87.09\% for recall, 83.90\% for F1, and	83.81\% for AUC, improving the results of textual-only embedding by 0.88\%, 0.12\%, 2.17\%, 1.14\%, and 0.92\%, respectively.
The improvement is reasonable as incorporating both
node type and node textual information can provide the model with more valuable node features for training.
However, we also find that the node type may lead to a negative impact on some scenarios, such as an average decrease of 0.53\% on five metrics for TreeLSTM with AST representation.
We think the possible reason for this phenomenon is the utilization of a simple linear layer to concatenate the representation of the textual embedding and type embedding.
Such a straightforward concatenation method hinders the extraction of a deep connection between the two representations for real-world software bugs, especially when applied to the challenging patch correctness assessment task.

\subsection{Potential of Pre-trained Code Models}
\begin{table}[t]
  \centering
  \caption{Comparison results of pre-trained code models}
    \begin{tabular}{c|ccccc|c}
    \toprule
    Models & Accuracy & Precision & Recall & F1    & AUC   & Average \\
    \midrule
    CodeBERT & 83.22\% & 82.48\% & 83.48\% & 82.93\% & 87.80\% & 83.98\% \\
    CodeT5 & 82.96\% & 81.86\% & 83.67\% & 82.69\% & 86.88\% & 83.61\% \\
    \midrule
    Transformer & 82.48\% & 78.97\% & 87.27\% & 82.87\% & 82.60\% & 82.84\% \\
    \bottomrule
    \end{tabular}%
  \label{tab:plms}%
\end{table}%

\textbf{Objective.}
Recently, pre-trained code models (PCMs) have achieved impressive results in software engineering tasks, such as program repair and test generation~\cite{zhang2026survey}. 
Since the strong performance of these models may come not only from code representations but also from large-scale pre-training and stronger model capacity, we did not include them in the main experiments, whose goal is to study the impact of code representations under a more controlled setting. 
In this section, we further examine the potential of PCMs for predicting patch correctness.

\textbf{Design.}
We select two representative PCMs, \ie the encoder-only CodeBERT and the encoder-decoder CodeT5. 
As both models are pre-trained on code sequences, we represent the input code as token sequences and feed them into the models. 
For comparison, we also include Transformer, the best-performing non-pre-trained sequence-based model in our study.

\textbf{Results.}
Table~\ref{tab:plms} shows the results.
Overall, both PCMs achieve competitive performance, indicating their potential to learn patch semantics and assess patch correctness. 
Among the three models, CodeBERT achieves the best overall performance, with 83.22\% accuracy, 82.48\% precision, 83.48\% recall, 82.93\% F1, and 87.80\% AUC. 
Compared with Transformer, CodeBERT improves accuracy, precision, F1, and AUC by 0.74\%, 3.51\%, 0.06\%, and 5.20\%, respectively, while yielding a lower recall.
CodeT5 also performs competitively, achieving an average score of 83.61\%, which is 0.77\% higher than Transformer. 
These results suggest that PCMs can be effective for APCA and may offer stronger discriminative ability than non-pre-trained sequence models. 
However, the improvements are still modest. 
More importantly, PCM-based and representation-oriented approaches are largely orthogonal: the former emphasizes stronger model architectures and pre-training, whereas the latter focuses on better encoding patch semantics. 
Therefore, an important direction for future work is to combine advanced PLMs with more effective code representations to assess patch correctness.


\subsection{Threats to Validity}

The first threat to validity is the selection of baselines.
In Section~\ref{sec:baseline}, we consider four APCA techniques as baselines to evaluate the effectiveness of four types of representations.
We may fail to consider other techniques (mentioned in Section~\ref{sec:related_work}) due to the rapidly growing community and the constant emergence of APCA approaches.
However, our selected baselines represent state-of-the-art (\eg the traditional dynamic-based PATCH-SIM) and are widely-adopted in almost all of previous studies~\cite{lin2022context,tian2020evaluating}.
Besides, these selected baselines cover various categories, such as ODS~\cite{ye2021automated} with hand-crafted features, Tian~\etal~\cite{tian2020evaluating} with code sequences, and CACHE~\cite{lin2022context} with ASTs.
Furthermore, our work mainly focuses on empirical evaluations of the effectiveness of different representations.
Thus, we believe the baselines have little impact on our results.

The second threat to validity is the evaluation benchmark.
In our experiment, we focus on the most popular Defects4J in the APR and APCA literature involving 835 bugs from 17 open-source projects.
It is unclear to what degree the results in our work can be generalized to other benchmarks.
Despite other datasets available~\cite{lin2022context}, they are quite easy to identify due to simple bug types, such as 99.9\% AUC~\cite{zhang2024appt}, making it difficult to assess true capabilities.
To address this threat, we collect 2,274 patches generated by 39 existing APR tools, to the best of our knowledge, which is the largest patch set of Defects4J in the literature.
It should be noted that, with the continuous release of newer APR tools on Defects4J, we observed that their contributions of unique patches (\ie those not already generated by earlier tools) have become increasingly limited.
In fact, some of their patches overlap with patches already included in our dataset. 
For this reason, while Table 1 does not explicitly list these tools, their outcomes are partially covered, and our dataset remains both comprehensive and the largest available.
Besides, we mitigate the potential bias by using multiple evaluation metrics to exhaustively assess the involved techniques.
In the future, we plan to continuously update the dataset by incorporating newer APR tools, such as Tare~\cite{zhu2023tare} and Recoder~\cite{zhu2021syntax}, to further strengthen the completeness of our benchmark.

The third threat to validity lies in the evaluation setting.
First, following most previous studies~\cite{lin2022context,tian2020evaluating}, we evaluate the performance of APCA approaches within a cross-validation setting, which is the common practice in the APCA community.
It is possible that patches generated by different APR tools for the same bug are split between the training and testing sets. 
However, in a real-world scenario, it would be impossible to use the label from one patch to predict the other for the same bug.
Second, there exist more potential code representations and models that are excluded in our experiments due to page limit.
Third, we focus on filtering overfitting patches following most prior work and the results may not be generalized to correct patch identification.
To address this threat, we construct a large benchmark with 2,274 patches from 39 APR tools, consider 15 representations and 11 models, and select seven baselines from different categories.
Besides, considering the scale of our empirical evaluation involving nearly 600 trained models, \ie (45 models in RQ1 + one model in RQ2 + 22 models in RQ3 + 51 models in Section~\ref{sec:node}) $\times$ 5-fold cross-validation, the current results are enough to demonstrate the performance of code representations in patch correctness prediction.
In the future, researchers could further explore the experiments under more potential scenarios, such as the cross-project evaluation and correct patch identification settings, and more representation-classifier combinations.

\subsection{Implication and Guidelines}
We summarize the following practical guidelines for future studies on patch correctness assessment.

\textit{Code representation is closely tied to the utilized model.}
The pipeline of code representation-based APCA can typically be divided into two parts: code representation for feature extraction and the corresponding model for binary prediction. 
The key research question lies in how to appropriately represent patched code snippets and determine the model architecture that can effectively learn the distribution of overfitting patches over correct patches.
As discussed in Section~\ref{sec:rq1}, we find that one representation may vary in performance with different models and no single representation consistently achieves the optimal results, such as the accuracy of CFG ranging from 75.59\% to 83.15\% with three GNN models.
In the future, when designing code representations, researchers are required to consider how to select more suitable models for effective source code representation and patch correctness learning.

\textit{Exploring more advanced representation learning is promising.}
As discussed in Section~\ref{sec:baseline}, four types of code representation achieve better or comparable performance against previous state-of-the-art APCA approaches, such as an accuracy of 83.73\% with the graph-based representation.
Our work serves as the first attempt to empirically explore the benefits of code representation in the APCA community with some basic representations and models, such as TF-IDF and LSTM.
Despite that, the preliminary results demonstrate the impressive performance in learning code semantics and predicting patch correctness.
Thus, we recommend that future work be conducted in the following two directions.
First, on top of existing foundational representations, it is crucial to design specific patch-oriented representation-based approaches that consider the characteristics of patches.
Second, given that a large body of code-related tasks has benefited from recent large language models~\cite{tian2020evaluating,zhang2024appt}, it is promising to leverage LLMs' powerful code-understanding capabilities to reason about patch correctness, rather than relying on basic models in our work.

\textit{Combining different types of code representation is challenging.}
As discussed in Section~\ref{sec:fusion}, it is promising to combine two types of representations, such as an accuracy improvement of 13.58\% when integrating the sequence-based representation into the heuristic-based representation.
However, it is challenging to combine more types of representations in our preliminary experiments, such as an accuracy of only 82.21\% for the fusion of the heuristic-based, sequence-based, and tree-based representations.
Future research efforts can be leveraged to explore how to better integrate a multitude of representations from two aspects.
The first approach involves designing advanced fusion strategies to combine representation vectors, such as addition, multiplication, and random interleaving, while the second approach focuses on employing more powerful fusion models, such as emerging multi-modal models.

\section{Related Work}
\label{sec:related_work}

Our work is mainly related to automated program repair, patch correctness assessment, and code representation.

\subsection{Automated Program Repair}
\label{sec:re_apr}
Existing APR approaches can be categorized into four main categories according to the hypotheses to generate patches.
\textbf{Heuristic-based APR} generates a large number of patch candidates based on syntactic program modifications and employs heuristics to explore the search space of patch generation~\cite{le2012genprog, martinez2016astor}.
For example, GenProg~\cite{le2012genprog} uses genetic programming to search for a patch that passes all provided test cases and SimFix~\cite{jiang2018shaping} searches for code snippets from code change operations from existing patches and similar code snippets within the buggy project.
\textbf{Constraint-based APR} transforms the patch generation into a constraint-solving problem and uses a solver to obtain a feasible solution.
For example, ACS~\cite{xiong2017precise} generates precise patches by ranking the ingredients for condition synthesis and is considered one of the representative approaches.
\textbf{Template-based APR} generates patches by employing pre-defined fix patterns to mutate buggy code snippets with the retrieved donor code \cite{koyuncu2020fixminer, liu2019tbar, liu2019avatar}.
For example, TBar~\cite{liu2019tbar} summarizes a variety of fix patterns from the literature and GAMMA~\cite{zhang2023gamma} is a successor of TBar by retrieving the donor code with mask prediction on top of LLMs.
\textbf{Learning-based APR}~\cite{tufano2019empirical, lutellier2020coconut, jiang2021cure} treats the APR problem as a neural machine translation task with the advance of DL models.
For example, Recoder~\cite{zhu2021syntax} predicts edits to generate syntactically-correct patches with a syntax-guided edit decoder and CIRCLE~\cite{yuan2022circle} can fix multilingual bugs with a single repair model based on continual learning and T5.
Furthermore, recent studies have explored the use of LLMs for program repair~\cite{xia2025demystifying,zhang2024fixing,applis2025unified}, which can be viewed as a further extension of learning-based APR. 
Since this direction is not the main focus of this paper, we do not discuss it in detail; a comprehensive overview is available in the survey~\cite{zhang2026survey}.

\subsection{Automated Patch Correctness Assessment}
\label{sec:re_apca}
Existing APCA techniques can be divided into three categories based on whether dynamic execution or machine learning is used, including static-based, dynamic-based and learning-based APCA, summarized as follows.

\textbf{Static-based APCA.}
Such techniques usually adopt static analysis tools to analyze the buggy program, extract some designed static features (such as context and operations), and then determine the correctness of patches based on these features.
For example, ssFix~\cite{xin2017ssfix} utilizes token-based syntax representation to generate patches with a higher probability of correctness.
Anti-patterns~\cite{tan2016anti} collects a total of 86 software bugs from 12 real-world projects and identifies a set of erroneous program modification patterns. 
S3~\cite{le2017s3} utilizes six syntactic features and semantic features to measure the distance between a candidate patch and the buggy code snippet. 
If the distance exceeds a pre-set threshold, the corresponding patch is identified as overfitting.

\textbf{Dynamic-based APCA.}
Such techniques mainly identify the correctness of patches based on the results of test case execution or the program's execution path. 
For example, Daikon~\cite{yang2020daikon} utilizes program invariants and oracle information to explore differences between overfitting and correct patches. 
For a patch, if its inferred invariant is identical to the true invariant, it is considered correct. If there exists a difference, the patch is considered overfitting.
Besides, DiffTGen~\cite{xin2017difftgen} identifies overfitting patches by measuring syntax differences between the patched and buggy programs by using an external test generator to generate new test cases and comparing the output of a patch on these test cases with that of the correct program on the same test cases. 
PATCH-SIM~\cite{xiong2018identifying} predicts patch correctness based on the assumption that the execution traces of passing tests on the buggy and patched programs may be similar, and the execution traces of failing tests on these two programs may differ.

\textbf{Learning-based APCA.}
Such techniques attempt to predict whether a plausible patch is correct or not based on machine learning techniques~\cite{zhou2024leveraging}.
For example, ODS~\cite{ye2021automated} first extracts 202 code features at the abstract syntax tree level and trains the classifier to predict which ones are correct patches.
Tian~\etal~\cite{tian2020evaluating} propose to utilize representation learning techniques to generate an embedding for buggy and fixed code snippets, and then input them into supervised learning classifiers to predict patch correctness.
Besides, CACHE~\cite{lin2022context} is a context-aware APCA technique by considering both code context and AST structural information.
Le~\etal~\cite{le2023invalidator} propose INVALIDATOR by utilizing program invariants and pre-trained models to perform semantic and syntactic reasoning for patch correctness.
Recently, Zhang~\etal~\cite{zhang2024appt} propose APPT, a pre-trained model-based APCA approach by jointly fine-tuning BERT and classifiers.

\subsection{Code Representation in SE}
\label{sec:re_representation}
The SE community has seen some studies to summarize and investigate code representation in various code-related tasks.
For example, Siow~\etal~\cite{siow2022learning} conduct an empirical study to evaluate four categories of code representation techniques across three tasks, including code classification, vulnerability detection, and clone detection.
Zhang~\etal~\cite{zhang2023survey} summarize the existing learning-based APR studies into three categories according to code representation, including sequence-based, tree-based, and graph-based approaches.
Similarly, Namavar~\cite{namavar2022controlled} invesgiate the impact of different code representations in learning-based APR approaches.
Utkin~\etal~\cite{utkin2022evaluating} conduct a preliminary empirical study to explore the impact of different AST parsers on the performance of Code2Seq and TreeLSTM on the task of method name prediction.
Sun~\etal~\cite{sun2023abstract} conduct a comprehensive empirical study to explore the effectiveness of AST representation on three tasks, including clone detection, code search, and code summarization.
Recently, Silva~\cite{silva2024repairllama}~\etal propose RepairLLaMA by combining realistic repair-specific code representations with parameter-efficient fine-tuning.
Unlike the above studies, we are the first to explore the performance of various code representation approaches in predicting patch correctness.

\section{Conclusion and Future Work}
\label{sec:conclusion}


In this paper, we perform the first exploration to empirically investigate the feasibility of different code representations on automated patch correctness assessment.
Specifically, we conduct a large-scale study to analyze the effectiveness and limitations of patch prediction approaches supported by code representation, involving  15 features from four categories, 11 classifiers, and more than 500 trained models.
The results demonstrate that the graph-based representation can consistently outperform other representations, \eg an accuracy of 83.95\% for DDG with GGNN.
We further demonstrate representation approaches achieve comparable or better performance against state-of-the-art approaches, such as CAHCE.
We also explore the potential of fusing different types of code representation and discuss the impact of different node embeddings, such as textual and type node information.
Lastly, our findings reveal practical guidelines for future representation-based patch correctness assessment studies.
Overall, our work demonstrates the promising future and challenges of using code representation in reasoning about patch correctness and helping developers effectively exploit off-the-shelf APR tools in practice.
In the future, we will consider more powerful models and design code representations tailored for predicting patch correctness.
We also hope to introduce advanced strategies to fuse different types of code representations.

\section*{Acknowledgments}
This work is supported partially by the National Natural Science Foundation of China (61932012, 62141215, 62372228).

\bibliographystyle{ACM-Reference-Format}
\bibliography{reference}

\end{document}